\documentclass[manuscript,screen]{acmart}
\AtBeginDocument{%
  }

\usepackage{float}
\usepackage{tabularx}
\usepackage{amsmath}
\usepackage{amsfonts}
\usepackage{algorithmic}
\usepackage[ruled,vlined, linesnumbered]{algorithm2e}
\usepackage{multirow}
\usepackage{graphicx}
\usepackage[export]{adjustbox}
\usepackage{caption}
\usepackage[strict]{changepage}
\usepackage{rotating}
\usepackage[labelformat=simple]{subcaption}

\usepackage{array}
\usepackage{algorithm2e}
\usepackage{makecell}
\usepackage{booktabs}
\usepackage{pdflscape}
\usepackage{tablefootnote}
\usepackage{threeparttable}
\usepackage{tcolorbox}
\usepackage{hyperref}
\usepackage{soul}
\usepackage{enumitem}
\usepackage{tcolorbox}
\usepackage{tablefootnote}
\usepackage{ulem}
\usepackage{eqparbox}
\usepackage{arydshln}
\usepackage{multirow}
\usepackage{adjustbox}
\usepackage[table]{xcolor}

\newcommand{\PreserveBackslash}[1]{\let\temp=\\#1\let\\=\temp}
\newcolumntype{C}[1]{>{\PreserveBackslash\centering}m{#1}}
\newcolumntype{R}[1]{>{\PreserveBackslash\raggedleft}m{#1}}
\newcolumntype{L}[1]{>{\PreserveBackslash\raggedright}m{#1}}

\newcommand\Tstrut{\rule{0pt}{2.5ex}}       
\newcommand\Bstrut{\rule[-1.3ex]{0pt}{0pt}} 
\newcommand\TBstrut{\Tstrut\Bstrut}         

\begin{document}

\title{Experimental Analysis of Productive Interaction Strategy with ChatGPT:\\ User Study on Function and Project-level Code Generation Tasks}

\author{Sangwon Hyun}
\email{dr.sangwon.hyun@gmail.com}
\orcid{0000-0003-1885-3792}
\affiliation{%
  \institution{Adelaide University}
  \city{Adelaide}
  \state{SA}
  \country{Australia}
}

\author{Hyunjun Kim}
\email{hyunjun1121@kaist.ac.kr}
\affiliation{%
  \institution{Korea Advanced Institute of Science and Technology}
  \city{Daejeon}
  \country{Republic of Korea}
}
\author{Jinhyuk Jang}
\email{jhjangbot@kaist.ac.kr}
\affiliation{%
  \institution{Korea Advanced Institute of Science and Technology}
  \city{Daejeon}
  \country{Republic of Korea}
}
\author{Hyojin Choi}
\email{hyojinchoi@kaist.ac.kr}
\affiliation{%
  \institution{Korea Advanced Institute of Science and Technology}
  \city{Daejeon}
  \country{Republic of Korea}
}
\author{Muhammad Ali Babar}
\affiliation{%
  \institution{Adelaide University}
  \city{Adelaide}
  \state{SA}
  \country{Australia}
}
\email{ali.babar@adelaide.edu.au}

\renewcommand{\shortauthors}{Hyun et al.}

\begin{abstract}
The application of Large Language Models (LLMs) is growing in the productive completion of Software Engineering tasks. Yet, studies investigating productive prompting techniques often employed a limited problem space, focusing primarily on well-known prompting patterns and targeting function-level SE practices. We identify significant gaps in real-world workflows that involve complexities beyond class-level (e.g., multi-class dependencies) and different features that can impact Human-LLM Interaction (HLI) processes in code generation. To address these issues, we designed an experiment to comprehensively analyze HLI features related to code generation productivity. Our study presents two project-level benchmark tasks that extend beyond function-level evaluations. We conducted a user study with 36 participants from diverse backgrounds, asking them to solve the assigned tasks by interacting with the GPT assistant using specific prompting patterns. We also examined participants’ experiences and behavioral features during interactions by analyzing screen recordings and GPT chat logs. Our empirical investigation, based on statistical guidance revealed (1) that three out of 15 HLI features emerged as consistently supported factors for productivity; (2) five primary guidelines for enhancing productivity for HLI processes; and (3) a taxonomy of 29 runtime and logic errors that can occur during HLI processes, along with suggested mitigation plans. 
\end{abstract}

\ccsdesc[500]{Human-centered computing~Empirical studies in HCI}
\ccsdesc[300]{Software and its engineering~Software development process management}


\keywords{Human-LLM Interaction, Code Generation Productivity, User Study}


\maketitle

\section{Introduction}

It is widely recognized that developers in various domains utilize Large Language Models (LLMs), particularly ChatGPT, to complete different Software Engineering (SE) tasks~\cite{liang2024large}. 
While the practicality and productivity with and without LLMs have been highlighted~\cite{ziegler2024measuring}, the challenge of how developers can effectively interact with LLM code assistants for optimal productivity remains understudied.

Human-LLM interactions can be facilitated through prompts that guide users in designing instructions and providing context for the model~\cite{kojima2022large}. Various prompting techniques have introduced example-based methods (e.g., Zero-Shot or Few-Shot), which are effective for general LLM applications in information retrieval and reasoning~\cite{kojima2022large,brown2020language}. These prompt patterns are actively utilized across various SE tasks, including code and test case generation~\cite{lin2025soen,mathews2024test,yuan2023no}, bug reproduction~\cite{kang2023large}, and program repair~\cite{ruiz2024novel}. However, most studies assume that these prompting techniques can yield optimal outcomes, which could lead to potential risks of redundant iterations and diminished developer acceptance of the outcomes, indicating lower productivity in SE tasks~\cite{ziegler2024measuring}.
For example, developers might repeatedly refine their prompts to generate more accurate test suites or codes, only to discover minor differences each time. This would trap the developer in an inefficient spinning wheel and undermine potential productivity gains.

To address these concerns, we investigated LLM-driven code and test case generation research~\cite{lin2025soen,mathews2024test,yuan2023no,piya2024llm4tdd,du2024evaluating}, empirical analysis studies using LLMs to solve SE tasks~\cite{rahe2025programming,barke2023grounded,fakhoury2024llm, liang2024usability, sergeyuk2025practice,weisz2024enterprise,stray2025humanai}, and studies suggesting a prompt programming and interaction structures ~\cite{liang2025promptprogramming,treude2025taxonomy}. Our findings reveal that (1) the majority of the research remains centered on function-level or narrowly scoped class-level tasks that exhibit limited dependency and diversity; (2) most studies have not taken into account comprehensive Human-LLM Interaction (HLI) features, such as user experience on using LLMs and users' prompt design behaviors, in their analyses; (3) the investigation into errors arising during HLI processes, along with the suggestion of mitigation strategies, has remained understudied.

The first challenge is that most existing studies have focused on function-level code generation and bug fixing. 
Du et al.~\cite{du2024evaluating} addressed class-level code generation by introducing a class-level code generation benchmark, ClassEval, and noted that LLM performance in this task is limited due to dependencies between functions and classes. This suggests that the complexities of real-world engineering tasks, arising from dependencies, inheritance, and API integrations among various functions, classes, and third-party providers, cannot be adequately addressed by conventionally used function-level SE tasks. Furthermore, many function-level benchmarks rely on a limited set of source datasets, highlighting diversity issues in the codes used for assessment~\cite{paul2024benchmarks}. Therefore, class or project-level engineering tasks must be considered to evaluate the effectiveness of prompting techniques in addressing the reality gaps.


Next, most existing studies concentrated on analyzing prompt patterns, such as Zero-Shot or Few-Shot, in different SE tasks. Recent studies have proposed usability-, workflow-, or context-management-driven interaction approaches~\cite{liang2024usability, sergeyuk2025practice, weisz2024enterprise, stray2025humanai} in addition to the prompt patterns. However, numerous other features, including user background, process-level strategies, and time distribution, can influence HLI processes in code generation. For example, users' approaches to defining prompt context, or the types of prompt context used across test cases and design descriptions, should be explored to establish practical guidelines for interacting with LLM code assistants. However, current research has not thoroughly examined the comprehensive features of HLI processes in its analyses.

Lastly, the in-depth analysis of errors and their root causes that occurred during the HLI processes is limited. Although Liu et al.~\cite{liu2024refining} examined various static and runtime errors generated by LLMs through a scalable empirical analysis, the study primarily focused on errors arising during function-level code generation tasks and did not consider user-perspective features as root causes. Therefore, strategies for mitigating errors induced by specific user behavioral patterns, multi-class dependency issues, and runtime or logic errors require further investigation.

Taken together, prior work leaves three unresolved gaps: limited evidence beyond function-level tasks, limited analysis of interaction behavior beyond the nominal prompt pattern, and limited investigation of how user-side and model-side behaviors shape downstream errors. To address these gaps, we designed a controlled empirical study involving 36 participants who solved function-level and dependency-rich project-level tasks using predefined prompting patterns. The main contribution of the paper is an evidence-backed characterization of interaction behavior under these conditions, with statistical analysis used as a supportive feature-screening step.

Our empirical assessment of code generation productivity, measured by the number of test cases passed within the allocated time, together with supportive exploratory analysis of interaction traces, points to three main findings: (1) among the HLI features considered in this study, prompt pattern, time allocation between initial implementation and debugging, and context-curation behavior emerged as the features most consistently associated with productivity under the study conditions; (2) in particular, higher-performing participants more often used Few-Shot prompting, allocated relatively more time to debugging than to initial implementation, and combined CopyPaste with Manual Formulating when curating context from test resources; and (3) accross the collected traces, we identified 29 categories of runtime and logic errors attributable to both user-side and GPT-side faults, highlighting recurring uncertainties in the implementation and debugging of multi-class code generation. These findings were interpreted from the empirically grounded observations in this study.


This study presents the following key contributions:
\begin{itemize}
    \item Newly introduced two multi-class code generation benchmarks to compare various prompting strategies.
    \item Analyzed best practices for productive prompting strategies for code generation, encompassing various features in the HLI processes.
    \item Developed a taxonomy for runtime and logic errors, including their root causes and suggestions for minimizing uncertainties during interaction with GPTs.
\end{itemize}

The paper is organized as follows: Sections 2 and 3 explain the works related to this study and the background on HLI features used in our experiment. In Section 4, we elucidate the experiment design. Sections 5 and 6 describe the empirical analysis results with discussion points. Finally, Section 7 concludes the study with directions for future work.

\section{Related Work}
\label{sec.related}

Existing literature relevant to this paper can be organized into three research themes. First, existing studies have evaluated LLMs for automated software engineering tasks such as code generation, unit test generation, class-level coding, and TDD-inspired workflows~\cite{lin2025soen,mathews2024test,yuan2023no,piya2024llm4tdd,du2024evaluating}. Second, empirical studies have investigated how developers and students actually use AI code assistants, including interaction patterns, usability frictions, workflow-specific usage, and perceived productivity~\cite{rahe2025programming,barke2023grounded,fakhoury2024llm, liang2024usability, sergeyuk2025practice,weisz2024enterprise,stray2025humanai}. Third, recent work suggests that prompts and developer-AI interaction structures should be treated as first-class software-development artifacts~\cite{liang2025promptprogramming,treude2025taxonomy}. Lastly, broader software-engineering and HCI research emphasizes that developer-AI collaboration should be interpreted through human-centered lenses such as multidimensional productivity, developer experience, trust calibration, and responsible adoption~\cite{forsgren2021space,noda2023devex,hoda2023augmentedagile,copenhagen2024manifesto,ahn2024trust,sergeyuk2026hax, ziegler2024measuring}.


\textbf{Automated SE Tasks by LLMs.} The studies on automated code or test generation have compared various prompting methods and SE processes to complete function-level tasks. Lin et al.~\cite{lin2025soen} developed multi-role LLM agents, such as Architecture Designer and Software Engineer, and assessed different software development processes utilizing these agents on HumanEval. Mathew et al.~\cite{mathews2024test} investigate how integrating Test-Driven Development (TDD) principles into LLM-based code generation can enhance the correctness of the generated code, primarily focusing on the effectiveness of test cases. Piya et al.~\cite{piya2024llm4tdd} also introduced a TDD framework that incrementally incorporates manually designed test cases to enhance code correctness. Yuan et al.~\cite{yuan2023no} conducted a comprehensive evaluation of unit test generation by LLMs and proposed a method for improving output quality. Du et al.~\cite{du2024evaluating} established a fundamental structure and examples for a class-level benchmark dataset for SE tasks, revealing that existing LLMs struggle to perform class-level SE tasks accurately. However, while these studies predominantly focused on function-level SE practices, Du et al.~\cite{du2024evaluating} also emphasized developing a benchmark for class-level practices. They employed Zero-Shot and One-shot patterns and did not explore different prompting methods in experiments.

\textbf{Empirical Analysis on SE Tasks.} 
Several recent studies have conducted empirical user research on function-level tasks. 
Rahe et al.~\cite{rahe2025programming} examined how programming students interact with LLMs to solve coding exercises. Their study with 37 students noted that students tend to use LLMs to seek knowledge about general concepts or to generate solutions directly. Barke et al.~\cite{barke2023grounded} analyzed the empirical usage patterns of 20 users with Copilot while solving customized experimental task sets. Fakhoury et al.~\cite{fakhoury2024llm} designed a TDD workflow process for developers and validated its effectiveness with 15 developers. Liu et al.~\cite{liu2024refining} conducted a scalable examination on GPT-generated codes for function-level tasks and defined a taxonomy of static and semantic errors. Most recent work broadens this Human-LLM Interactions by showing usability challenges of AI programming assistants~\cite{liang2024usability}, workflow-and activity-specific usage and non-usage patterns across software development process~\cite{sergeyuk2025practice}, enterprise evidence on perceived productivity on developer experience~\cite{weisz2024enterprise}, and mixed-method findings on how developers combine IDE-integrated assistants such as Copilot in organizational settings~\cite{stray2025humanai}.

\textbf{Prompt Programming and Interaction Structure.} A third category of the research has treated prompts and interaction structures as development artifacts. A recent study by Liang et al. suggests that it is effective to treat the prompts as programs too and to characterize prompt programming as a distinct software-development activity with its own mental models, iteration patterns, and tool needs~\cite{liang2025promptprogramming}. Treude et al.~\cite{treude2025taxonomy} organized developer-AI collaboration into multiple interaction types across the software life cycle.  

\textbf{Human-Centered Productivity, Collaboration, and Trust.} The last theme mainly suggests that developer-AI collaboration should be evaluated with human-centered constructs. The SPACE framework shows that developer productivity is multidimensional rather than reducible to a single output metric~\cite{forsgren2021space}, while DevEx emphasizes the lived developer experience and friction points that shape productivity~\cite{noda2023devex}. Hoda et al.~\cite{hoda2023augmentedagile} proposed a human-centered AI-assisted software management approach, which further suggests that AI adoption in software engineering should augment rather than displace human judgment, while Copenhagen et al.~\cite{copenhagen2024manifesto} reinforced a position for human-centered generative AI in software engineering. Recent studies also connected AI-assisted development to perceived productivity in practice~\cite{ziegler2024measuring, weisz2024enterprise}, trust calibration process~\cite{ahn2024trust}, and the emerging literature on human-AI experience inside IDEs~\cite{sergeyuk2026hax}.

Across these themes, we have identified that extant studies have not comprehensively considered the features defined in our study, such as user background characteristics, prompt patterns, context-curation behavior, debugging strategy, and observed error outcomes in both function-level and class-level SE tasks. Most studies provide evidence on either benchmark-centric and function-level approaches, or on perceived productivity without trace-based analysis of how developers actually interact with assistants while solving interdependent coding tasks. Our study addresses this gap by combining controlled function-level and project-level tasks with screen recordings, GPT chat logs, survey data, and exploratory analysis of HLI features through statistical methods.

\section{Background}
\label{sec.background}

\subsection{Human-LLM Interaction Features}

In formulating HLIs for code generation practices, we defined three main attributes: User, Model, and Interaction by extending the 5W1H HLI catalogue format~\cite{gao2024taxonomy}. Then, we investigated concrete features for each attribute that can affect the engineering productivity with LLM code assistants.

Firstly, the \textbf{User} attribute abstracts users' backgrounds and experiences in handling assigned tasks and specifying prompts. For instance, the programming and industry experience features implicitly indicate users' skill levels and confidence in general engineering and industry project tasks. The LLM usage experience feature can reflect users' prompt engineering skills, while the dependency feature indicates the degree of trust in their prompting abilities and reliance on outcomes. The algorithmic experience represents the users' familiarity with algorithm-solving tasks, as most function-level tasks and class-level sub-tasks involve algorithm-solving challenges~\cite{chen2021evaluating,du2023classeval}. We observed that existing studies did not account for user-oriented features in their empirical and quantitative analyses of software engineering tasks. However, we incorporated the user attribute into our analytical features because their backgrounds and experiences may shape the user's overall decision-making regarding interaction strategy, ultimately affecting productivity in completing the assigned tasks.

\begin{table}[]
\centering
\caption{Overall Attributes and Features of HLIs on SE Tasks} \label{tab:HLI}
\begin{tabular*}{0.66\textwidth}{ll}
\hline \hline \multicolumn{2}{c}{\TBstrut \textbf{Human-LLM Interaction Features for Software Engineering Tasks}}                                                                                                                         \\ \hline
\multicolumn{2}{c}{\TBstrut User Perspective}                                                                                                                                                                               \\ \hline
\TBstrut Programming experiences                                                                      & 2-4 / 3-5 / 6-10 / 10+ years                                                                                        \\ \hdashline
\TBstrut Industry experiences                                                                         & 0: None / 1-2 years / 3-4 years / 5+ years 
\\ \hdashline
\begin{tabular}[c]{@{}l@{}} \Tstrut LLM usage experience \\ \Bstrut when conducting coding tasks\end{tabular} & \begin{tabular}[c]{@{}l@{}}0: Never / 25: rarely / 50: half the time /\\  75: most / 100: always\end{tabular}       \\ \hdashline
\begin{tabular}[c]{@{}l@{}} \Tstrut Dependency on LLMs \\ \Bstrut in final code outcomes\end{tabular}         & \begin{tabular}[c]{@{}l@{}}Applying 0-100\% of \\ codes generated from LLMs\end{tabular}                             \\ \hdashline
\begin{tabular}[c]{@{}l@{}}\Tstrut algorithm-solving experiences\\ \Bstrut in the past 2 years\end{tabular} & \begin{tabular}[c]{@{}l@{}} 0-5 / 6-10 / 11-15 / \\ 16-20 / 20+ problems\end{tabular}                            
\\ \hline
\multicolumn{2}{c}{\TBstrut Model Perspective}                                                                                                                                                                                \\ \hline
\TBstrut License                                                                                      & Free / Paid                                                                                                         \\ \hdashline
\TBstrut Interaction interface                                                                        & WebUI / IDE / API                                                                                                   \\ \hline
\multicolumn{2}{c}{\TBstrut Interaction Perspective}                                                                                                                                                                             \\ \hline
\TBstrut Interaction process                                                                 & Waterfall, Test-Driven Development, etc      \\ \hdashline    
Prompting pattern                                                                            & \begin{tabular}[c]{@{}l@{}}\Tstrut Few-Shot, CoT, Reflection, \\ \Bstrut Alternative Approaches, etc\end{tabular}              \\ \hdashline
\TBstrut Prompt design method                                                                         & \begin{tabular}[c]{@{}l@{}} \Tstrut Copy Paste / Formulate / Using LLMs \\ on prompting / \Bstrut Pre-existing template\end{tabular} \\ \hdashline 
\TBstrut Context Artifact                                                                                        & \begin{tabular}[c]{@{}l@{}} \Tstrut  Requirement specification, Test cases, \\ Architecture, Skeleton codes, etc. \Bstrut \end{tabular}   

\\ \hline \hline
\end{tabular*}
\vspace{-15px}
\end{table}

Next, the \textbf{Model} attribute indicates the ontological features of various LLM types, including their accessibility through licensing policies and interaction interfaces, which may influence user experiences in addressing engineering tasks or the quality of outcomes. We employed both the Free and Paid features in our experiment to compare the contributions of the feature with other user-side and interaction-side features in terms of code generation productivity. However, for the interface, we asked all the participants to use the same WebUI to unify other environment settings.

The \textbf{Interaction Method and Process} attributes are the primary goals of analysis addressed by existing studies~\cite{du2024evaluating,lin2025soen,mathews2024test}. We defined these attributes to instantiate the conceptual ``How to interact with LLMs concretely" feature~\cite{gao2024taxonomy} in the context of code generation tasks. The interaction process outlines specific steps and approaches for conducting HLIs on software engineering tasks, such as following a waterfall development process or a test-driven development approach. The interaction method involves techniques for designing prompts, considering existing pattern templates and various design methods, including copying and pasting specification documents and custom formulation, as well as the types of artifacts utilized in curating the context. We explain the prompting patterns in depth in the next Section and the coverage of the features in Section~\ref{sec.disc}.

\subsection{Prompting Pattern Templates}

Prompting patterns among the other HLI features have been extensively studied by existing research~\cite{kojima2022large,brown2020language,wei2022chain,liu2022design,arora2022ask}. Among these studies, White et al.~\cite{white2023prompt} categorized prompting strategies and designed a catalog of prompting pattern templates. Based on the templates, we identified two well-known patterns: Few-Shot and Chain-of-Thought (i.e., Cognitive Verifier), as well as two improvement-driven prompting patterns: Reflection and Alternative Approach patterns, which are designed to conduct the reasoning process with LLMs. These Reflection and Alternative Approach patterns facilitate reasoning about self-repairing and feedback loops, which are known to improve code generation qualities~\cite{liu2024refining}. The templates for these patterns are systematically designed to represent the conventional use cases of each template, based on the actual prompts used in existing studies. All the prompt patterns are available in our experiment repository\footnote{https://github.com/abalon1210/Multi-Class-LLM-CodeGen-Benchmark}.



\begin{figure}[t]
    \centering
    \includegraphics[width=0.65\textwidth, trim = 0cm 0.4cm 0cm 0cm,clip]{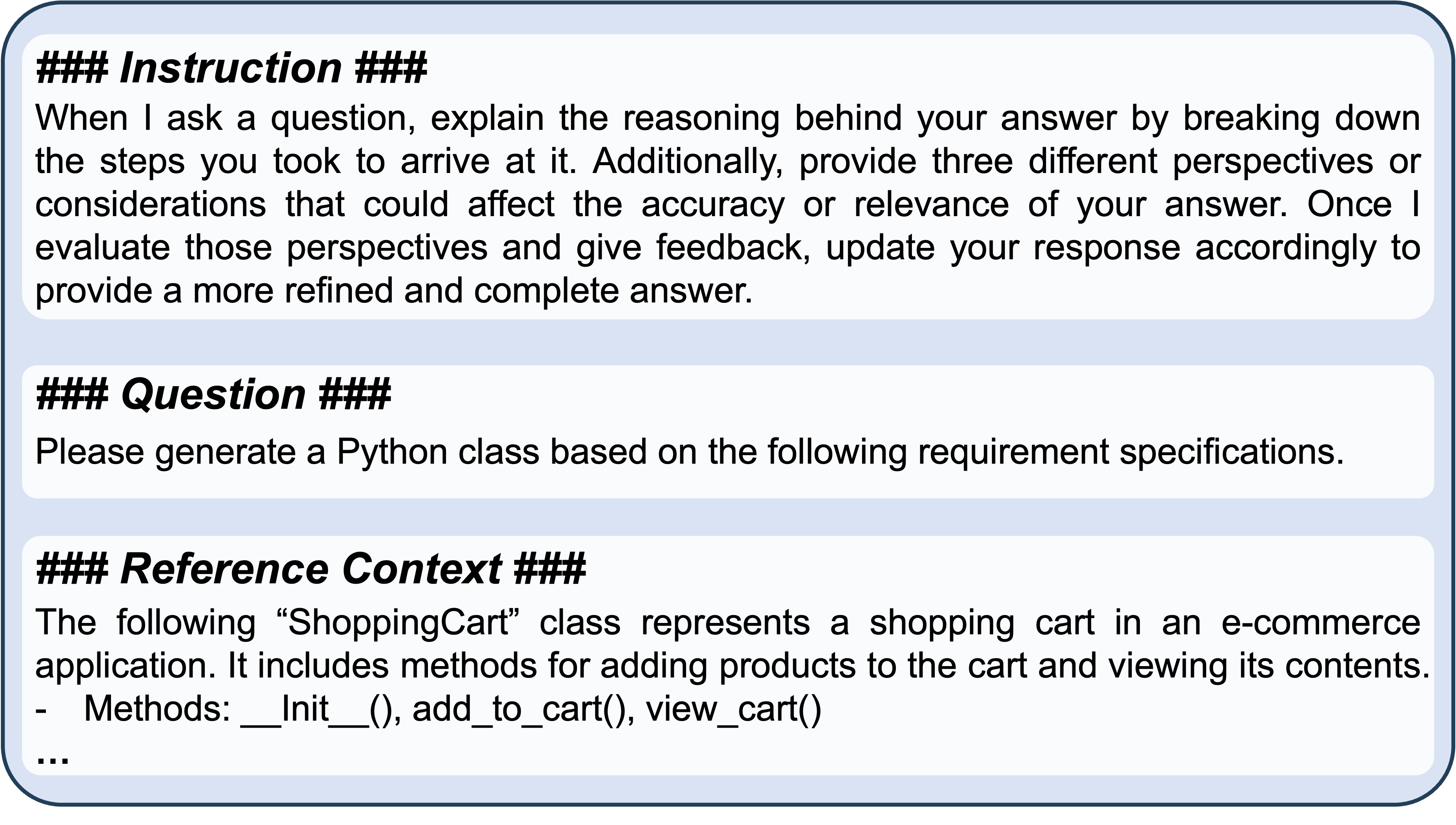}
    \caption{Example prompt template for the Reflection pattern} \label{fig:background}
    \vspace{-20px}
\end{figure}

First, the \textbf{Reflection} pattern intends to get GPTs to automatically explain the reasoning behind the answers they give to the users' questions. This can help better understand how GPTs process the input and the assumptions they make in their outputs. 
We expect the Reflection pattern to aid in improving outcome codes by explaining the underlying logic to users, allowing them to verify correctness and debug any issues. Fig.~\ref{fig:background} illustrates the prompt template example for the reflection pattern used in our experiment, containing the pre-defined Instruction part to explain the Reflection pattern, and the Question and Reference Context parts where users can add any request and context information.

The \textbf{Alternative Approach} pattern aims to ensure that GPTs provide multiple methods for accomplishing a task. This prompting pattern is motivated by users becoming fixated on pursuing only a single approach. 
This pattern aims to help users recognize alternative methods they may not have considered, enabling them to choose the most suitable option. We anticipate that this prompt pattern can offer various approaches for developing or fixing code, allowing users to select the best solution.

The \textbf{Cognitive Verifier} (a.k.a Chain-of-Thought) pattern enables GPTs to subdivide an original question into smaller questions that can ultimately yield a better answer to the original one. This pattern employs a divide-and-conquer strategy based on the premise that GPTs can reason more effectively when a question is divided, and the answers to these smaller questions can be combined to form a solution to the original question. Although this pattern can determine the number of questions GPTs should ask users, it means that users must answer those questions to receive appropriate answers from GPTs.

Lastly, we defined the well-known \textbf{Few-Shot} template by adding two pre-defined request-and-answer examples in the Example part below the three-part structure described in Fig.~\ref{fig:background}. Since optimizing prompt design is not within the scope of our study, we focused on establishing a conventional template with two fixed examples for each task in this Few-Shot pattern. The number of examples was determined based on existing research~\cite{lin2025soen, mathews2024test, yuan2023no, kang2023large, ruiz2024novel}. For each task, we included one specific example of generating partial classes or functions alongside one general example. To minimize the variance caused by the prompt pattern templates, we asked participants not to modify the order and number of examples in the Example part when using this pattern. Still, we allowed users to add any context in the Reference Context part. One example of a Few-Shot pattern template usage is explained in the Appendix.

\section{Experiment Design}

This section outlines our user-participating experiment process, function and project-level coding tasks, and concrete research questions with corresponding productivity measures adopted in our experiment.

\subsection{Research Questions}

The following Research Questions (RQs) are defined to explore the most productive strategy for interacting with GPTs on code generation at the function-level and project-level tasks:
\begin{itemize}[leftmargin=10px]
    \item RQ1. Which HLI features show the strongest and the most consistent associations with code generation productivity under the study condition?
    \item RQ2. What best practice guidelines can be established for strategic interactions with GPTs to enhance productivity?
    \begin{itemize}[leftmargin=15px]
        \item RQ2-1. Which prompting patterns yield the highest productivity for both function-level and project-level software engineering tasks?
        \item RQ2-2. What features distinguish the Best quartile group from the other groups?
    \end{itemize}
    \item RQ3. What errors occur during HLI processes for code generation?
\end{itemize}

\textbf{RQ1} investigates which HLI features show the strongest and most consistent associations with code generation productivity under the study conditions. We use the Test Passed Rate (TPR) as our primary productivity measure. Because this study models a relatively broad set of HLI features with a modest sample size, we frame RQ1 as an exploratory feature-screening analysis rather than a confirmatory causal test. To that end, we apply a regularized and resampling-based statistical procedure to identify candidate user-side, model-side, and interaction-side features that recur across complementary analyses.

\textbf{RQ2} focuses on investigating key aspects of HLI strategies and providing best practice guidelines for using ChatGPT to enhance code generation productivity. \textbf{RQ2-1} aims to examine the relationship between the prompt patterns and their impact on TPR productivity. In \textbf{RQ2-2}, we conducted a group-wise comparison of different features between the Best quartile group and the other quartile groups. We analyzed context curation methods, such as context design methods (e.g., Copy-Paste, Manual-Formulating, etc.) and different types of contextual artifacts (e.g., design document, test cases, etc.) that can affect the productivity of interactions. We identified the first quartile group of TPR as our best practice group. We thoroughly analyzed the screen recordings and GPT communication logs of participants in this group, comparing them with those of participants in other groups. The findings from these research questions provide clear guidelines for HLI best practices for multi-class code generation with GPT assistance.

\textbf{RQ3} concentrates on the empirical analysis of errors detected in the user study to categorize the types of errors, identify the root causes of errors, and suggest mitigation strategies for identified errors. We first define a taxonomy of HLI errors and outline the types, frequencies, and symptoms of the 791 identified errors. Then, we analyzed the root causes of the errors and identified whether the errors are user-side or GPT-side faults by thoroughly reviewing the participants' outcome codes, chat logs, and screen recordings. We believe these findings offer valuable insights into error handling in code generation, particularly in the context of using GPT models.

\subsection{Overall Experiment Process}

\begin{figure*}[t!]
    \includegraphics[width=0.98\textwidth, trim = 0cm 0.7cm 0cm 0cm,clip]{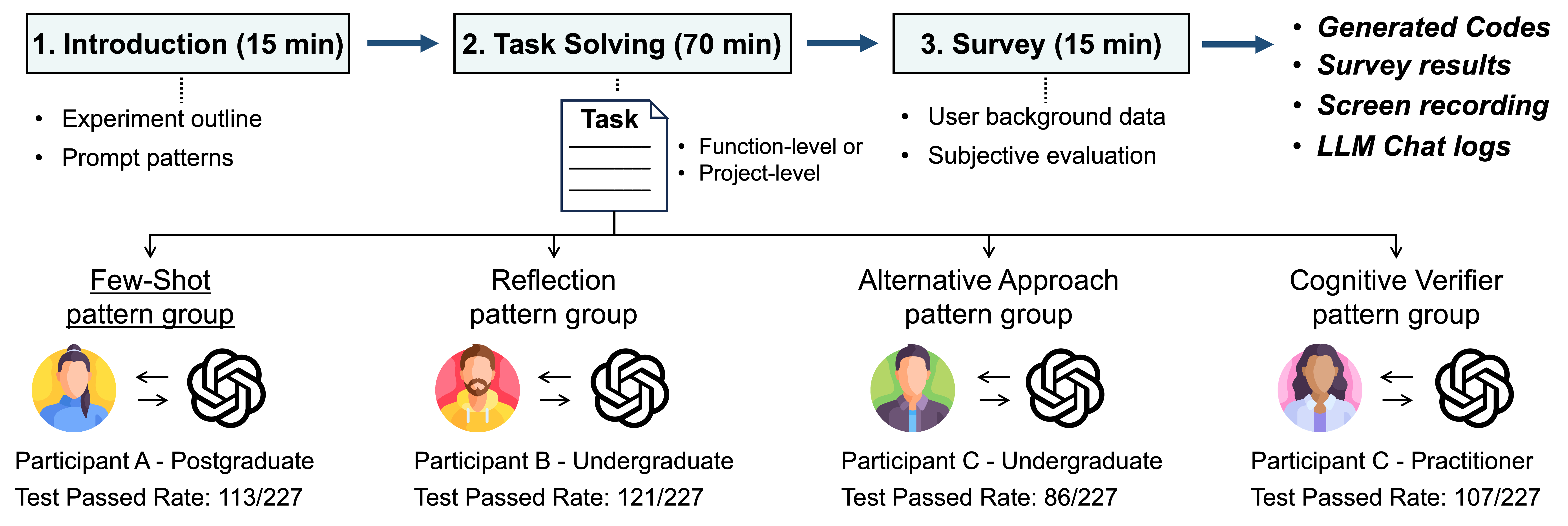}
    \caption{ Overall experimental process and example outcomes of interactive problem solving on project-level tasks} \label{fig:overall}
    \vspace{-15px}
\end{figure*}

Our experiment aims to identify the most productive prompting strategy for completing code generation tasks. We have designed problem-solving tasks at the function and project levels, distributed the design documents with skeletons to users, and asked them to generate code within a specified time. 

Fig.~\ref{fig:overall} illustrates the \textbf{Experimental Process}, outlining the first 15 minutes to introduce the experiment to participants and explain the assigned prompt patterns, 70 minutes for actual task solving, and the last 15 minutes for surveying the in-depth background and subjective evaluations on prompting patterns and their design methods. The experiment was conducted online and offline, depending on the participants' circumstances. The introduction session aims to explain the experiment process, rules, and tasks with prompt patterns, through prepared materials. 

\begin{table}[t!]
\centering
\caption{Study Design Summary Table}
\label{tab:study-design}

\begin{tabular}{p{4cm} p{10cm}}
\hline \hline
\textbf{Element} & \textbf{Summary} \\
\hline
\TBstrut Recruitment & 
Participants were recruited through flyers, LinkedIn, and Upwork. Thirty-six participants who satisfied the Python-experience criterion were included. \\ \midrule

\TBstrut Task level & 
Participants were assigned to function-level or project-level tasks, with 12 participants in the function-level condition and 24 in the project-level condition. \\ \midrule

\TBstrut Experience balancing & 
Participants were distributed as evenly as possible according to Python experience. For instance, the function-level group included 4 participants with 2--3 years, 4 with 4--5 years, and 4 with over 6 years of Python experience. \\ \midrule

\TBstrut Prompt pattern & 
Four prompt patterns were evaluated: Few-Shot, Reflection, Alternative, and Cognitive Verifier. Each pattern was assigned to 9 participants. \\ \midrule

\TBstrut GPT tier & 
Participants were divided into Free and Paid GPT tiers, with 18 participants in each tier. GPT tier was balanced within experimental groups where practical. \\ \midrule

\TBstrut Exposure & 
Each participant worked under one prompt-pattern condition and one task-level condition. \\ \midrule

\TBstrut Session structure & 
Each session included a 15-minute introduction, a 70-minute task-solving phase, and a 15-minute post-task survey. \\ \midrule

\TBstrut Provided artifacts & 
Participants received a design document, skeleton code, executable tests, and a prompt-template format. \\ \midrule

\TBstrut Compensation & 
Participants were paid USD 50 after submitting the required artifacts. \\ \midrule
\hline \hline
\end{tabular}
\vspace{-15px}
\end{table}

For the \textbf{Participant Recruitment}, we disseminated our experiment through various methods, like hands-on flyers and LinkedIn, and recruited 36 participants via the Upwork platform~\cite{Upwork}. Our recruitment criteria require more than two years of Python programming experience. We recruited participants globally with diverse backgrounds, including undergraduate and postgraduate students and practitioners with industry experience. We evenly distributed the participants into four comparison pattern groups based on their backgrounds and project logs on Upwork.

The \textbf{Task Solving} phase was executed by distributing the design documents, prompt pattern templates, and skeleton Jupyter Notebook files or problem links to participants. We assigned function-level tasks to 12 participants and project-level tasks to 24 participants.
Additionally, we strictly limited participants' manual coding only when GPTs could not generate the expected outcomes after three request attempts. This concentrates on interactions with users and GPT code assistants and their outputs while preventing situations where biased contributions from users or GPTs affect the completion of SE tasks. 
The detailed task and test suite design are specified in Section~\ref{sec.design.task}. 

For the \textbf{GPTs} that participants interacted with, we considered the Free-tier and subscription-tier (i.e., Paid) GPTs they primarily use in their regular practices. Based on the Expression of Interest questions on Upwork, we assigned participants Free or Paid models according to their preferences in this experiment. Overall, 18 participants utilized free-tier GPTs, such as GPT-3.5 or GPT-4o-mini, while the other half adopted the Paid 4o model. The concrete distributions of GPT models by tasks and prompt pattern groups are illustrated in Section~\ref{sec.expr.stats}.

The \textbf{Survey} aims to gather specified user background information and their subjective assessment of prompting strategies. We have designed 30 multiple-choice and text-entry questions that collect specific attributes that can influence the interaction strategy and processes between participants and GPTs, as detailed in Table~\ref{tab:HLI}.


The final question set evaluates participants' subjective views on their interaction strategies. We provided sample scales to assess participants' subjective productivity, accuracy, and efficiency of their prompting strategies, rated on a scale from None to Strongly. Additionally, we asked them to share specific experiences where GPTs produced outcomes with improved or diminished qualities. The last questions of our survey asked whether they would like to adopt these prompting strategies in their future studies and to share their comments on the experiments.

\subsection{Task Design: Function and Project-level} \label{sec.design.task}

Our study specified concrete design documents for function and project-level tasks. We thoroughly designed test cases aligned with the specifications and skeleton codes for convenient distribution and immediate task-solving execution.

\textbf{Function-level Tasks.} Several benchmark datasets, such as HumanEval~\cite{chen2021evaluating}, are available for assessing the code generation accuracy of GPTs. However, we decided not to utilize the problems in existing benchmarks because their tasks have been segmented, indicating that each task contains only a single requirement for code generation or fixing. Since our work focuses on analyzing interactions between GPTs and engineers, we determined that these segmented SE tasks could not entirely capture potential interactions, and additional processes would be necessary to curate or combine them. Moreover, there may be data leakage concerns where GPTs directly respond to answer codes.

Utilizing existing problems from popular coding challenge platforms, such as LeetCode~\cite{LeetCode} and HackerRank~\cite{HackerRank}, can offer an accessible range of coding tasks at various levels of difficulty~\cite{ziegler2024measuring,barke2023grounded,tian2023chatgpt}. For function-level tasks, we treated data leakage as a task-selection concern rather than a model-evaluation variable. Candidate problems from LeetCode and HackerRank were screened by querying multiple GPT models with only the public task description. We excluded tasks for which the models directly returned near-complete canonical solutions with minimal interaction and retained tasks that still required multi-step clarification, debugging, or refinement. The final set contains two Easy, two Medium, and one Hard problem and was intentionally chosen to require substantive interaction within the 70-minute session. The problem list is detailed in our repository\footnote{https://github.com/abalon1210/Multi-Class-LLM-CodeGen-Benchmark.}

\textbf{Project-level Tasks.} 
The E-Commerce App and Smart Home Gym tasks were designed as controlled dependency-rich proxies for conversational code-assistant use, not as full replicas of an industrial codebase. Their purpose is to expose interaction challenges that are largely absent from function-level tasks, including multi-class dependency management, state propagation across classes, incremental integration, and iterative debugging under time pressure. We therefore deliberately specified several classes, cross-class method dependencies, executable tests, and notebook-based scaffolding, while excluding unrelated concerns, such as UI implementation or external deployment infrastructure. We specified classes and methods rigorously, manually designed test cases based on the equivalent partitioning method~\cite{burnstein2006practical} that cover both general and edge cases, and included skeleton code in Jupyter Notebook to facilitate direct task solving and assessment.

\begin{figure*}[t!]
    \includegraphics[width=0.98\textwidth, trim = 0cm 0cm 0cm 0cm,clip]{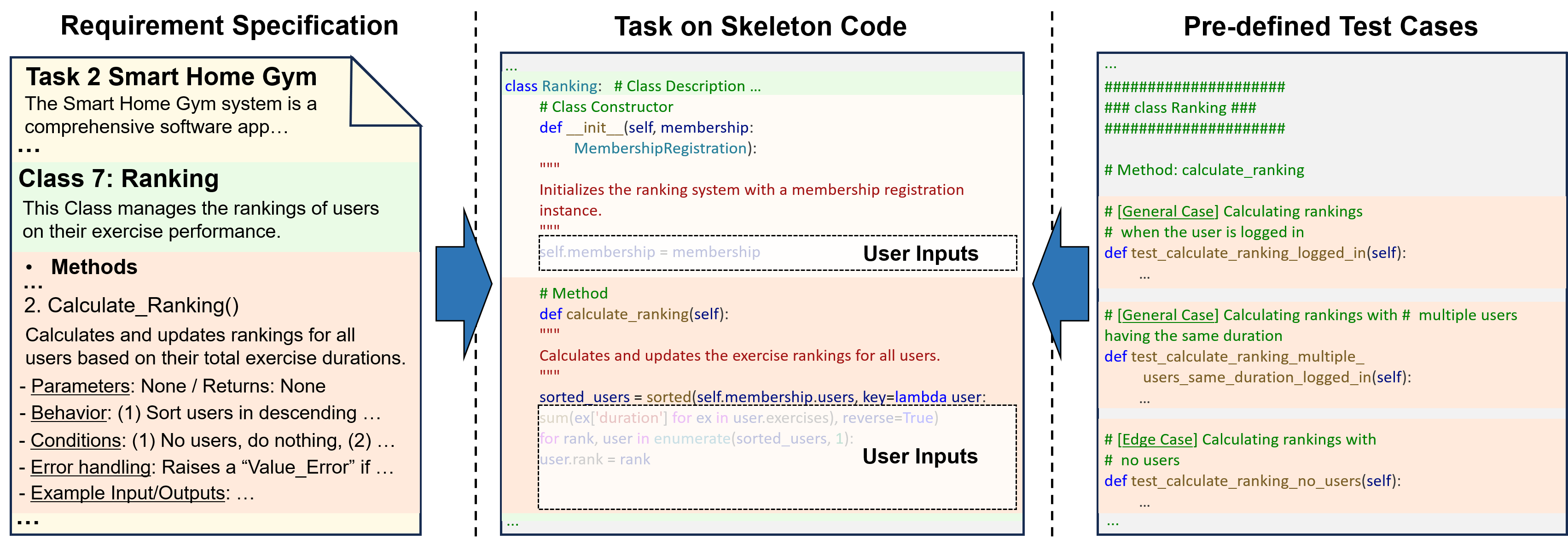}
    \caption{ Example design specification along with the corresponding skeleton code snippets and test cases} \label{fig:example}
\end{figure*}

Our project-level tasks consist of design document, skeleton code for task solving, and executable test cases for corresponding requirements. Examples of each component are illustrated in Fig.~\ref{fig:example}. First, our specification documents for each project-level task outline the project’s abstract and goals, provide an overview description of each class, and detail method specifications. In particular, the method specifications clearly state the essential features and guidelines for developing the correct functions, including parameter and return information, specific behaviors, special requirements, error-handling guidelines, and example inputs and outputs. We have designed 5 classes with 14 methods for the E-Commerce scenario task and 7 classes with 22 methods for the Smart Home Gym task. We did not include any UI-related requirements or implementations in the tasks.

The middle part of Fig.~\ref{fig:example} provides an example skeleton code for the Ranking class in the Smart Home Gym task. Following the class specification format proposed in ClassEval~\cite{du2024evaluating}, we designed the skeleton code to include descriptions of classes and methods, parameter details, and sections for user input. Additionally, the dependency examples between the Ranking and Membership classes are demonstrated in the two functions included in the skeleton code. We have concentrated on developing task components that can evaluate the capability of GPT's code generation in managing software code dependencies for project-level tasks.

Lastly, we designed a test suite for each project task by applying the equivalent partitioning method~\cite{burnstein2006practical}, considering all the requirements and their input spaces. We categorized the test cases into general and edge cases. The general cases aim to evaluate the expected functionalities of the developed classes and methods, while the edge cases focus more on assessing the implementation of special requirements (E.g., No-user case) and error-handling guidelines. In total, we designed 128 test cases for the E-Commerce App and 100 test cases for the Smart Home Gym task. We ensured that the test suite for each task achieved 100\% branch coverage in our sample codes. Additionally, we empirically checked that the test suites achieved an average line coverage of 91.85\%, excluding non-executable code, and an average branch coverage of 99.83\%, excluding redundant conditional branches, in the final code sets from participants.
The test cases are also included in the distributed Jupyter Notebook files, allowing participants to conveniently execute the tests to evaluate and improve the code generated by GPT assistants. Additionally, manual coding was permitted only after three unsuccessful GPT requests on the unresolved issue, so that the study could remain centered on the interaction process while still allowing bounded progress only when the model stalled. A more detailed project-level task example is described in the Appendix, and all the files are provided in the attached file.

For the project-level task experiment, the E-Commerce App served as Task 1 and the Smart Home Gym as Task 2. Task 2 became available only after participants achieved a passing score of 121/128 on Task 1. We defined this staged rule based on our internal testing and adapted it to emulate a workflow in which developers must stabilize one subsystem before moving to the next. Importantly, the project-level TPR was computed across all 228 test cases; therefore, participants who did not advance to Task 2 had zero solved tests for that second-stage task. Accordingly, Task 2 should be interpreted as a continuation of the project-level workflow rather than as an independently randomized comparison across all 24 participants.

\subsection{Productivity Measures for GPT Outcomes}
\label{sec.design.measure}

Software engineering productivity for human engineers has been extensively studied in previous research~\cite{boehm1987improving,scacchi1995understanding,mccabe1976complexity,ISO25002:2024}. These studies have highlighted the importance of considering various factors when assessing SE productivity, such as correctness, efficiency, maintainability, security, and subjective evaluations, including developer satisfaction. For example, correctness has been measured by test-pass or requirement-completion rates, while the time-to-completion metric can be used to measure efficiency. 
Recent studies utilized the SE productivity measure to evaluate the code generation productivity of GPTs~\cite{ziegler2024measuring,barke2023grounded,coutinho2024role}. These studies concentrated on the correctness aspect of GPT code generation; thus, they primarily employed the Pass@K~\cite{chen2021evaluating} or Test Passed Rate methods. In our experiment, we assessed the productivity of GPT code assistants by first applying the Test Case Passed Rate over 70 minutes and then using time-to-completion if all test cases were satisfied.
 
\section{Experiment Results}
\label{sec.expr}

\subsection{Statistics of Experiments} \label{sec.expr.stats}

\begin{figure}[t]
    \centering
    \includegraphics[width=0.65\textwidth, trim = 0.5cm 0.5cm 0.3cm 0.3cm,clip]{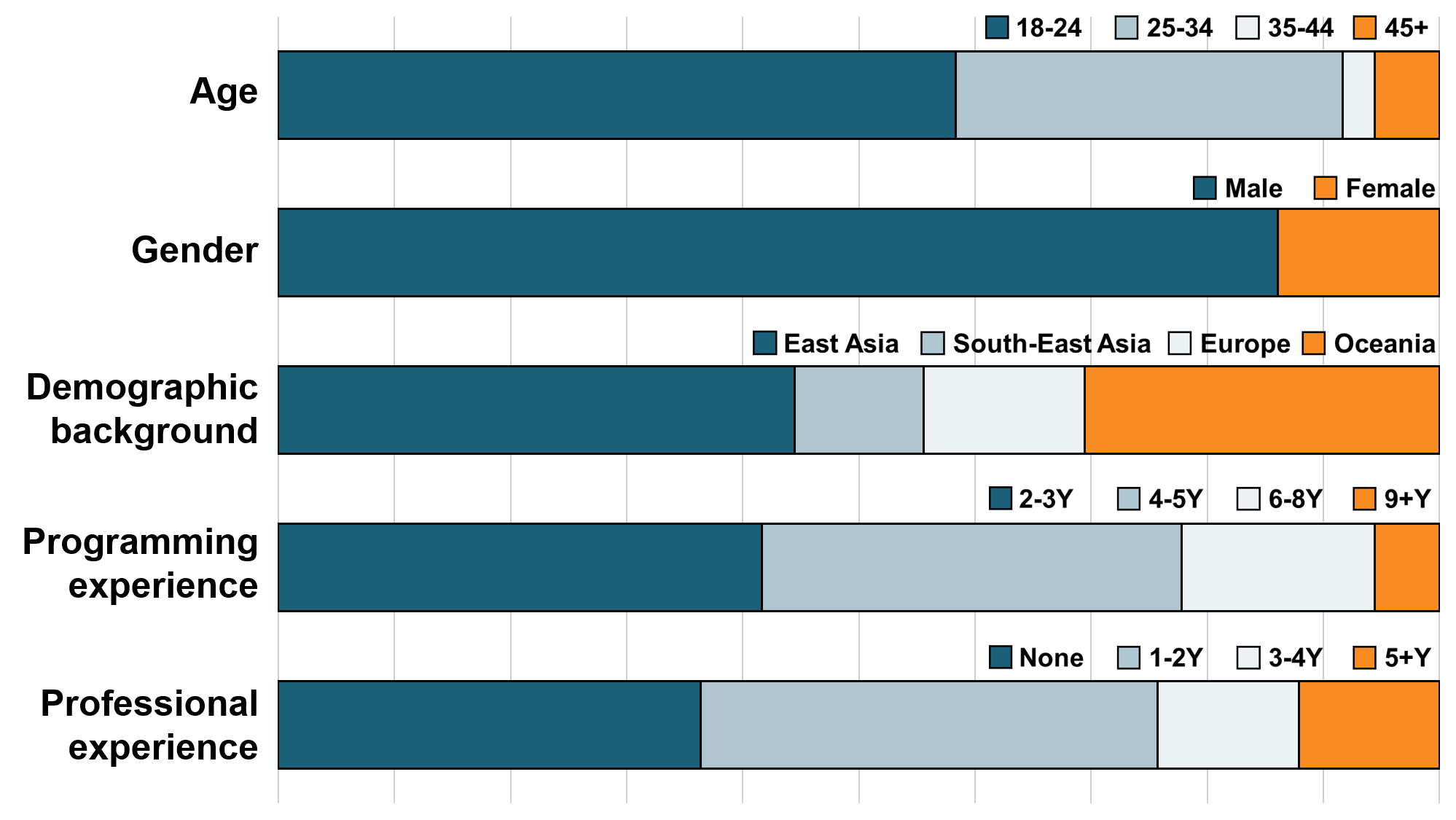}
    \caption{Overview of participant background statistics} \label{fig:stat_background}
    \vspace{-10px}
\end{figure}

\begin{figure}[t]
    \centering
    \includegraphics[width=0.65\textwidth, trim = 0.5cm 0.5cm 0.5cm 0.3cm,clip]{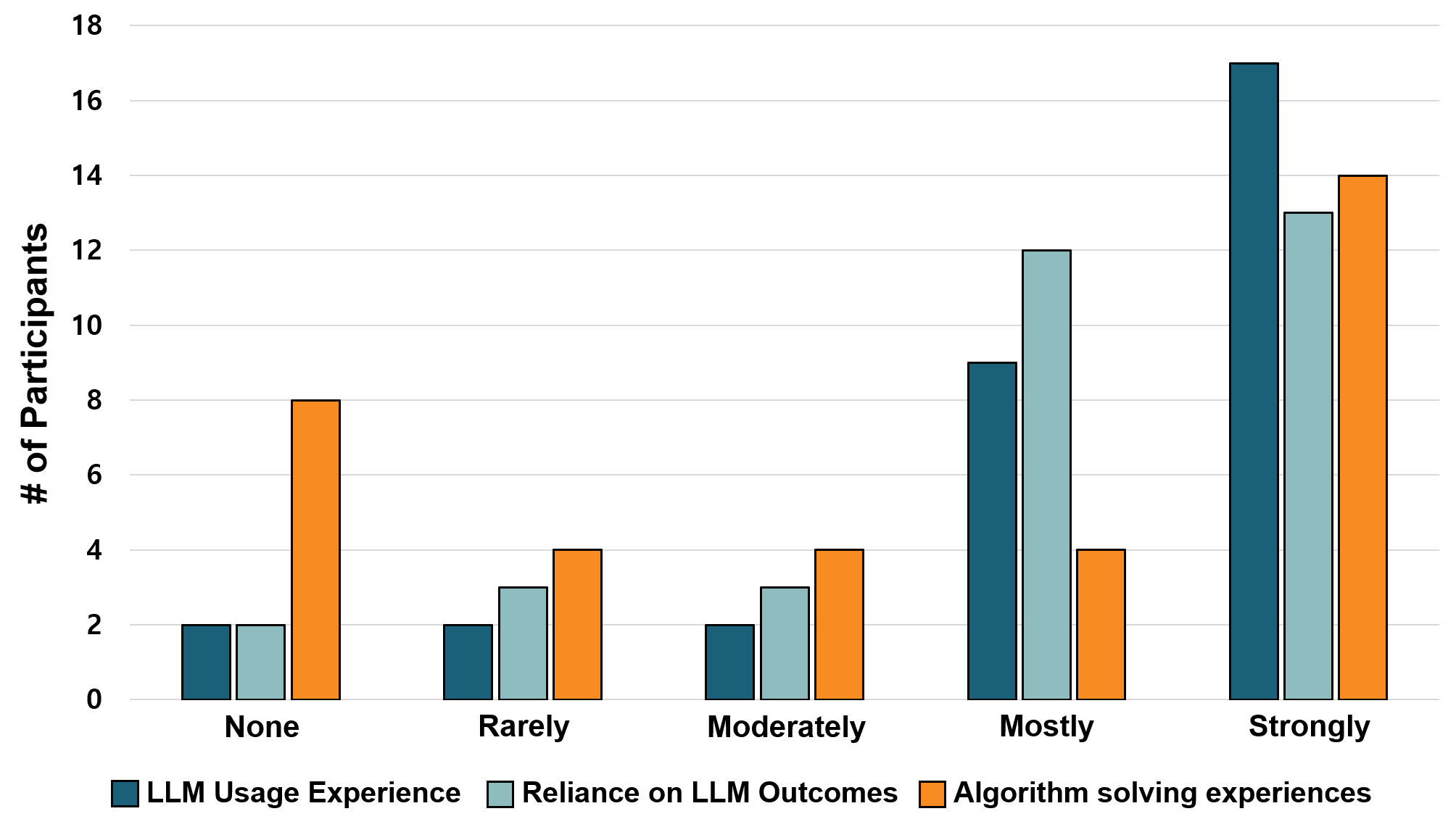}
    \caption{Overview of GPT usage and task-solving experience} \label{fig:stat_GPT}
    \vspace{-10px}
\end{figure}

Fig.~\ref{fig:stat_background} and \ref{fig:stat_GPT} illustrate the statistics of participants in our experiment. Fig.~\ref{fig:stat_background} presents a 100\% Stacked Bar Chart showing the backgrounds of the 36 participants. Among these, 21 participants (58\%) are in the age range of 18-24, 12 participants are aged 25-34 (33\%), one participant is aged 35-44, and two are aged 45 and older. A total of 31 male participants and 5 female participants participated in our experiment. Next, 16 participants were from East Asian backgrounds, 4 were from Southeast Asia, 5 were from Europe, and 11 were from Oceania. 15 participants have 2-3 years of Python programming experience, 13 have 4-5 years, while 9 participants have over 6 years of experience, including 2 with more than 9 years. In terms of professional experience, 12 participants have none, indicating they are undergraduate students, 14 have 1-2 years of experience, and 9 participants have more than 3 years of industry or research engineering experience, with 4 having over 5 years. Our experiment considered various user background factors and covered a diverse spectrum of users. 
The participant allocation was approximately balanced, not purely randomized, with programming experience, professional experience, and GPT tier considered during assignment across the four prompt-pattern groups. The exact between-subjects structure is summarized in Table~\ref{tab:study-design}.

We classified participants' status based on their academic and industrial experiences, as we deemed it more appropriate than categorizing them by their current status in our analysis. For example, one participant has over 9 years of industry experience and is a Ph.D. student, while another has 2 years of industry experience and is a 4th-year student. We utilized the categorized results to investigate whether academic or industry experience can serve as distinguishing features of optimal productivity in conducting software engineering tasks. To clarify the status of participants, our experiment consists of 18 undergraduate students, more than 2nd-year (3: Function-level/15: Project-level), 8 graduate students, including Master’s and Ph.D. programs (6: Function/2: Project), and 10 practitioners (3: Function/7: Project).

Fig.~\ref{fig:stat_GPT} illustrates the number of participants whose responses are categorized into five intensity levels based on their experience with GPT usage, reliance on GPT results, and experiences in task-solving for the last 2 years, including algorithm problems. For the GPT usage experience in real-world SE tasks, 26 participants stated that they mainly or strongly utilize GPTs. In comparison, only two participants indicated that they rarely use GPTs or not at all. In the reliance on code outcomes from GPTs, 25 participants answered they mostly or absolutely rely on GPT code outcomes. Lastly, for the task-solving experiences, the participants are nearly divided into two groups: familiar for the highest two groups, with 17, and non-familiar for the lowest two groups, with 12, based on their task-solving experiences over the past two years. We believe these background experiences may affect the decision on strategies of high-level task-solving processes, thereby contributing to the productivity of solving SE tasks with GPT assistants. 

Additionally, we assigned both Free and Paid GPTs to participants across four prompt pattern groups, considering function and project-level tasks. In total, 18 participants utilized Free-version GPTs, including models like 3.5 or 4o-mini, while the remaining 18 used Paid-version models such as 4o or o1. For every 9 participants in each group, we ensured that at least four were assigned to use both the Free and Paid versions. Furthermore, we assigned at least one Free and one Paid model to each function and project task subgroup. For instance, we allocated four Free models and five Paid models to the Few-Shot pattern groups, where two Free models and one Paid model were assigned to the Function-level task subgroup, and two Free and four Paid models were used for the Project-level group. Considering the varying circumstances of each participant regarding subscriptions to and use of GPT models, we distributed Free and Paid models across each categorical group in our experiment.

\subsection{RQ1: Exploratory Analysis of HLI Factors Associated with Code Generation Productivity} \label{sec.expr.rq1}
This RQ aims to analyze the overall contribution of HLI features defined in this experiment and obtain insights into which features most significantly impact the code generation productivity (i.e., Test Passed Rate (TPR) values) in HLI processes with GPTs. In total, we utilized 15 features in our comprehensive analysis, which included eight user-side features, one model-side feature, and six interaction-related features from 36 participants in the function and project-level tasks.

The analyzed features comprise eight user-side features, one model-side feature, and six interaction-related ones. The user-side features include age, current status, gender, programming experience, industrial programming experience, self-reported GPT usage experience, dependency on GPT outcomes, and algorithm-solving experience over the past two years. The model-side feature indicates whether the participant used a Free-tier or Paid-tier GPT model. The interaction-side features include prompting pattern, four context-curation features: Copy-Paste, Manual Formulation, Use of GPTs, and Utilization of External Templates, and the time-distribution strategy between initial implementation and debugging. The context-curation and time-distribution features were identified through a detailed review of participants' screen recordings and GPT chat logs.

Because the number of modeled features is large relative to the sample size, we treat the quantitative analysis in RQ1 as an exploratory feature-screening step rather than a confirmatory causal analysis. We first encoded categorical variables and standardized numeric variables. We then applied one-standard-error Elastic Net regularization to identify candidate features, followed by debiased estimation of intervals and p-values for the selected features, and bootstrap-based stability selection across 200 resamples. These procedures help reduce overfitting risk and highlight associations that recur under resampling, but they do not remove the inferential limitations imposed by the sample size. Accordingly, we interpret the resulting coefficients, intervals, and stability rates as supportive evidence for prioritizing candidate features for closer empirical inspection rather than definitive estimates of causal effect.

\begin{table}[]
\centering
\caption{Function-level Task Evaluation Results} \label{tab:RQ1_Overall}
\begin{tabular*}{0.66\linewidth}{lcccc}
\hline \hline
\Tstrut \textbf{Feature}                      & \textbf{$\boldsymbol{\beta}$-EN} & \textbf{$\boldsymbol{\beta}$-DL (95\% CI)}  & \textbf{p-value}& \textbf{Bootstrap} \\ \midrule
\begin{tabular}[c]{@{}l@{}} \textbf{Prompt Pattern:}\\ \textbf{Few-Shot}\end{tabular}             & \textbf{\underline{$+$0.058}}        & \begin{tabular}[c]{@{}c@{}}0.061\\ (+0.024 ... +0.098)\end{tabular}  & \textbf{\underline{0.002}}   &  \textbf{\underline{91\%}}                 \\ \midrule
\begin{tabular}[c]{@{}l@{}} \textbf{Time Distribution}\\ \textbf{Strategy}\end{tabular}           & \textbf{\underline{$+$0.034}}        & \begin{tabular}[c]{@{}c@{}}+0.036\\ (+0.007 ... +0.064)\end{tabular} & \textbf{\underline{0.014}}    & \textbf{\underline{92\%}}                 \\ \midrule
\begin{tabular}[c]{@{}l@{}} \textbf{algorithm-solving} \\ \textbf{Experience}\end{tabular} & \textbf{\underline{$-$0.033}}       & \begin{tabular}[c]{@{}c@{}}-0.030\\ (-0.061 ... -0.004)\end{tabular} & \textbf{\underline{0.028}}   & \textbf{\underline{88\%}}                 \\ \midrule
Paid Tier                    & $+$0.021        & \begin{tabular}[c]{@{}c@{}}+0.018\\ (-0.011 ... +0.043)\end{tabular} & 0.19        & 47\%                 \\ \midrule
\Bstrut Other 11 Features               & 0            & --                          & --         & $<$40\%        \\   \hline \hline
\end{tabular*}
\vspace{-10px}
\end{table}

Table~\ref{tab:RQ1_Overall} summarizes the results of the statistical screening analysis, focusing on the comparative contributions of each feature to the TPR values. The $\beta$-EN column reports the coefficients from the ElasticNet model trained on our experimental data. The third, $\beta$-DL, and fourth, p-value, columns indicate the 95\% confidence intervals and the significance of each feature inferred from the Debiased Lasso algorithm. The last column presents the bootstrap selection rates, which are used to estimate the confidence of the statistical outcomes. These indicators allow us to examine not only the direction of association, but also whether a candidate feature remains visible across complementary analysis.

The results presented in Table~\ref{tab:RQ1_Overall} indicate that three features emerged as the most consistently supported candidate factors in this analysis: prompting pattern, time-distribution strategy, and algorithm-solving experience. Among them, the prompt-pattern signal, particularly Few-Shot prompting, shows the clearest positive association with TPR and the strongest stability across the complementary analysis. The time-distribution strategy also shows a positive association, suggesting that participants who shifted relatively more effort from initial implementation to iterative debugging tended to achieve higher productivity under the study conditions. In contrast, the algorithm-solving experience shows a negative association in this dataset. We do not interpret this signal as evidence that stronger technical ability lowers productivity. Rather, our trace-based inspection suggests that it may reflect differences in interaction style, especially the tendency of some participants with stronger algorithmic backgrounds to rely more on manually formulated context during early development. We examine this behavior in more detail in Sections 5.3.2 and 5.4. 

The Paid GPT tier shows a positive direction, but weaker and less stable support than the three features above. We therefore treat the GPT tier as a secondary signal rather than a central conclusion of RQ1. The remaining 11 features do not show sufficient stability in this sample to justify a strong interpretation. Importantly, this does not imply that those features have no relationship with productivity; instead, it indicates that they did not emerge consistently enough in the present dataset to be prioritized in the subsequent empirical analysis. We use these signals as anchors for the more detailed task-level and trace-based analysis in the following sections.

\begin{tcolorbox} 
In RQ1, we found that prompt pattern, debugging-oriented time allocation, and interaction behavior related to context curation are the HLI features most consistently associated with code generation productivity.
\end{tcolorbox}

\begin{table}[]
\centering
\caption{Function-level Task Evaluation Results on Prompt Patterns} \label{tab:RQ2_Function}
\begin{tabular*}{0.68\linewidth}{cccccc}
\hline \hline
\multicolumn{2}{c}{\textbf{Metrics}}  & \textbf{Few-Shot}    & \textbf{Reflection}  & \begin{tabular}[c]{@{}c@{}} \Tstrut \textbf{Alternative} \\  \Bstrut \textbf{Approach}\end{tabular} & \begin{tabular}[c]{@{}c@{}} \Tstrut \textbf{Cognitive} \\ \Bstrut \textbf{Verifier}\end{tabular} \\ \hline \hline
 \Tstrut \multirow{3}{*}{\begin{tabular}[c]{@{}c@{}} Test \\  Passed \\ Rate  \end{tabular}}        & \Bstrut Min & \textbf{\underline{1.0}}         & 0.90        & 0.92                 & \textbf{\underline{1.0}}                \\
                                                                                       & \Bstrut Avg & \textbf{\underline{1.0}}         & 0.97        & 0.97                 & \textbf{\underline{1.0}}                \\
                                                                                       & \Bstrut Max & 1.0         & 1.0         & 1.0                  & 1.0                \\ \hline
\Tstrut \multirow{3}{*}{\begin{tabular}[c]{@{}c@{}}Subjective \\ Productivity\end{tabular}}    & \Bstrut Min & 80          & 80          & 80                   & 80                 \\
                                                                                       & \Bstrut Avg & \textbf{\underline{93.33}} & 86.67 & 86.67          & \textbf{\underline{93.33} }       \\
                                                                                       & \Bstrut Max & 100         & 100         & 100                  & 100                \\ \hline
\Tstrut \multirow{3}{*}{\begin{tabular}[c]{@{}c@{}}Subjective \\ Accuracy\end{tabular}}        & \Bstrut Min & 20          & 80          & 80                   & 60                 \\
                                                                                       & \Bstrut Avg & 66.67 & \textbf{\underline{93.33}} & 86.67          & 73.33        \\
                                                                                       & \Bstrut Max & 100         & 100         & 100                  & 100                \\ \hline
\Tstrut \multirow{3}{*}{\begin{tabular}[c]{@{}c@{}}Subjective \\ Efficiency\end{tabular}}      & \Bstrut Min & 40          & 80          & 60                   & 80                 \\
                                                                                       & \Bstrut Avg & 73.33 & 86.67 & 80                   & \textbf{\underline{93.33} }       \\
                                                                                       & \Bstrut Max & 100         & 100         & 100                  & 100                \\ \hline
\Tstrut \multirow{3}{*}{\begin{tabular}[c]{@{}c@{}}Future \\ Usage \\ Preference\end{tabular}} & \Bstrut Min & 80          & 80          & 80                   & 80                 \\
                                                                                       & \Bstrut Avg & \textbf{\underline{93.33}} & 86.67 & 86.67          & \textbf{\underline{93.33} }       \\
                                                                                       & \Bstrut Max & 100         & 100         & 100                  & 100                \\ \hline \hline
\end{tabular*}
\vspace{-10px}
\end{table}

\subsection{RQ2: Best Practice Guideline on HLIs for Code Generation} \label{sec.expr.rq2}
Based on our high-level findings regarding the key features that contribute to productive HLIs in code generation, we conducted an in-depth investigation to identify the best practices for interacting with GPTs for code generation tasks.

\subsubsection{RQ2-1: Analysis on Prompt Patterns} \label{sec.expr.rq2-1}

Tables \ref{tab:RQ2_Function} and \ref{tab:RQ2_Project} present the productivity calculation and subjective evaluation results of different prompt patterns for function and project tasks, respectively. While the TPR primarily serves as a comparison metric for different prompting patterns, the subjective evaluation results are also used as supplementary scores, indicating practical scores regarding productivity, accuracy, efficiency, and future usage preference.

\begin{table}[]
\caption{Project-level Task Evaluation Results} \label{tab:RQ2_Project}
\begin{tabular*}{0.68\linewidth}{cccccc}
\hline \hline
\multicolumn{2}{c}{\textbf{Metrics}}  & \textbf{Few-Shot}    & \textbf{Reflection}  & \begin{tabular}[c]{@{}c@{}} \Tstrut \textbf{Alternative} \\  \Bstrut \textbf{Approach}\end{tabular} & \begin{tabular}[c]{@{}c@{}} \Tstrut \textbf{Cognitive} \\ \Bstrut \textbf{Verifier}\end{tabular} \\ \hline \hline
 \Tstrut \multirow{3}{*}{\begin{tabular}[c]{@{}c@{}} Test \\  Passed \\ Rate  \end{tabular}}        & \Bstrut Min & \textbf{\underline{0.41}}     & 0.11       & 0.07                 & 0.16                \\
                                                                                       & \Bstrut Avg & \textbf{\underline{0.48}}     & 0.33       & 0.33                 & 0.32                \\
                                                                                       & \Bstrut Max & 0.52     & \textbf{\underline{0.53}}       & 0.48                 & 0.39                \\ \hline
\Tstrut \multirow{3}{*}{\begin{tabular}[c]{@{}c@{}}Subjective \\ Productivity\end{tabular}}    & \Bstrut Min & 5        & 29         & 0                    & 40 \\
                                                                                       & \Bstrut Avg & 62.17    & \textbf{\underline{66.17}}      & 53.33        & 65.17        \\
                                                                                       & \Bstrut Max & \textbf{\underline{100}}      & \textbf{\underline{100}}        & 91                   & 96      \\ \hline
\Tstrut \multirow{3}{*}{\begin{tabular}[c]{@{}c@{}}Subjective \\ Accuracy\end{tabular}}        & \Bstrut Min & 5        & 33         & 20                   & 50 \\
                                                                                       & \Bstrut Avg & 60.33    & 62.17      & 63.83                & \textbf{\underline{67.67}}  \\
                                                                                       & \Bstrut Max & 85       & \textbf{\underline{100}}        & \textbf{\underline{100}}                  & 83             \\ \hline
\Tstrut \multirow{3}{*}{\begin{tabular}[c]{@{}c@{}}Subjective \\ Efficiency\end{tabular}}      & \Bstrut Min & 50       & 61         & 40                   & 20   \\
                                                                                       & \Bstrut Avg & 70.67    & \textbf{\underline{74.17}}      & 72.83                & 59.67   \\
                                                                                       & \Bstrut Max & 92       & \textbf{\underline{98}}         & 95                   & 91    \\ \hline
\Tstrut \multirow{3}{*}{\begin{tabular}[c]{@{}c@{}}Future \\ Usage \\ Preference\end{tabular}} & \Bstrut Min & 30       & 21         & 20                   & 23    \\
                                                                                       & \Bstrut Avg & 59.00    & 51.50      & 54.33                & \textbf{\underline{65.67}}   \\
                                                                                       & \Bstrut Max & \textbf{\underline{85}}       & 83         & 81                   & 81        \\ \hline \hline
\end{tabular*}
\vspace{-10px}
\end{table}

\textbf{Function-level Task Analysis.} The experiment results on function-level tasks presented in Table~\ref{tab:RQ2_Function} show that 10 out of 12 participants completed all 72 test cases designed for the five tasks within 70 minutes. The remaining two participants also achieved TPRs of 90\% and 92\%. Additionally, we found that the unsolved test cases are not just from the Hard problem; they consist of one or two edge cases from multiple problems. This indicates that GPT code assistants can effectively provide solutions for function-level requirements, regardless of the difficulties, but have limitations in addressing all edge cases. Notably, participants employing Few-Shot and Cognitive Verifier patterns demonstrated slightly higher productivity than those using Reflection and Alternative Approach patterns. 

The subjective evaluations of each pattern also reveal results similar to those of the TPR. The Few-Shot and Cognitive Verifier patterns received the highest average scores in surveys on subjective productivity and future usage preferences. This indicates that participants who used the patterns experienced an improvement in productivity while solving function-level tasks and strongly preferred utilizing them in future practical scenarios. In the subjective accuracy survey, participants who used the Reflection pattern reported experiencing highly accurate code from GPTs. Conversely, the Cognitive Verifier received the highest efficiency ratings.

\textbf{Project-level Task Analysis.} On the other hand, the experimental results from project-level tasks, as detailed in Table~\ref{tab:RQ2_Project}, reveal the relative difficulties that participants experienced in solving tasks with GPT code assistants. Overall, the maximum TPR is 0.53, indicating that only about half of the project-level design requirements, 121 out of 228 test cases, were resolved within 70 minutes. The minimum TPR identified in the project-level task-solving case is 0.07, indicating that the participant and GPT assistant developed codes that passed only 16 test cases despite the participant communicating intensively with the GPT for over 70 minutes.

In project-level task results, participants using the Few-Shot and Reflection patterns achieved the highest average and maximum TPR values, respectively. Especially, the Reflection pattern recorded the highest scores in subjective evaluations of productivity, accuracy, and efficiency from the participants. The Cognitive Verifier and Few-Shot patterns obtained the top average and maximum scores for future preferences, respectively. 

\begin{figure}[t]
    \includegraphics[width=0.65\textwidth, trim = 0.5cm 0.4cm 0.5cm 0.3cm,clip]{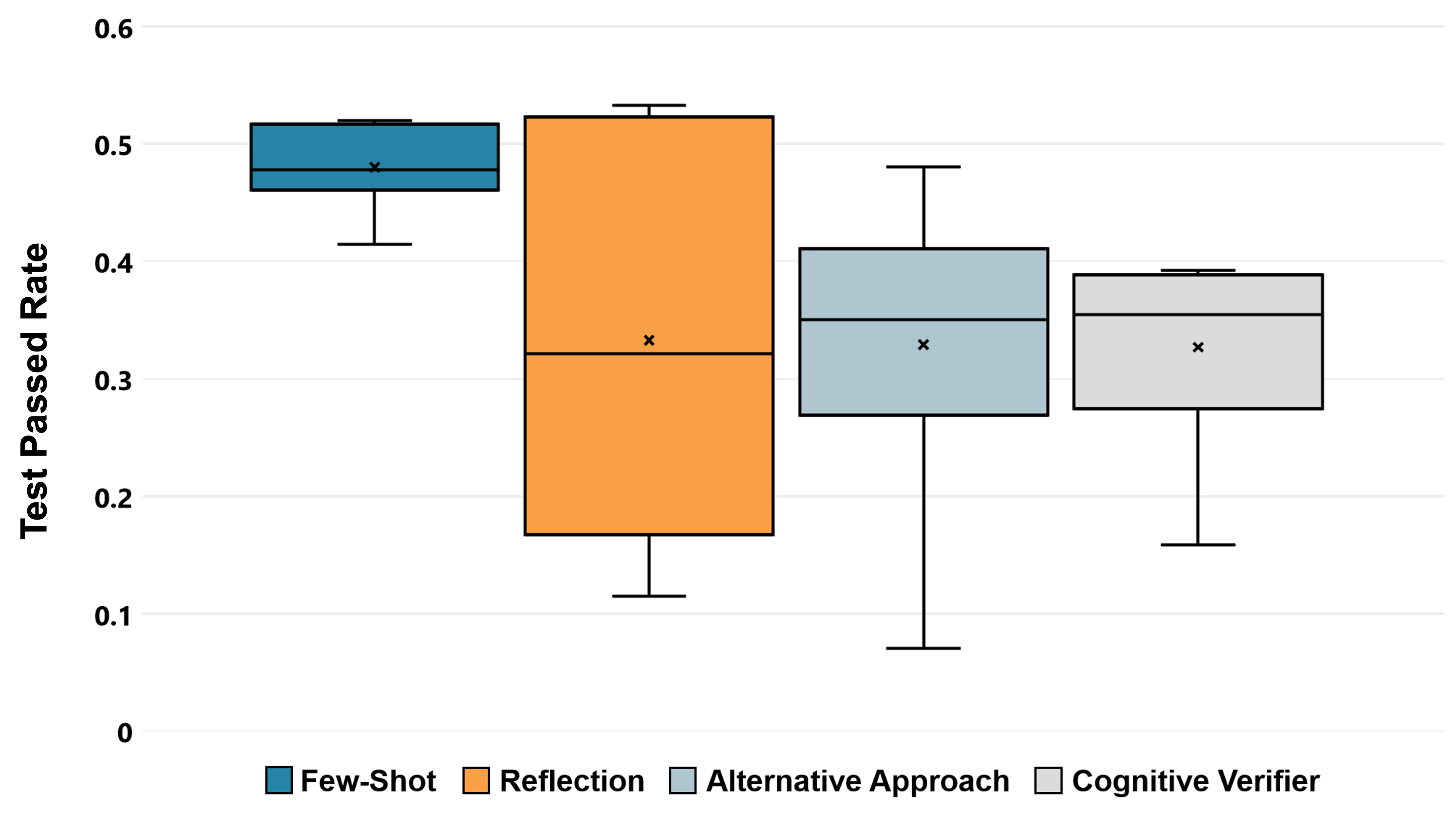}
    \caption{Productivity comparison results of prompt pattern groups in project-level SE tasks} \label{fig:RQ1}
    \vspace{-10px}
\end{figure}

Fig.~\ref{fig:RQ1} illustrates the distribution of TPR results for each prompt pattern group. The results indicate that while the Reflection pattern groups achieved the highest TPR score, the Few-Shot group exhibited significantly higher TPR distributions than the other groups. As a result, we performed the Wilcoxon Ranksum test~\cite{wilcoxon1992individual} between the Few-Shot group and the other groups. This analysis revealed that the Few-Shot group's TPR distributions have statistically significant differences compared to the Alternative Approach and Cognitive Verifier groups (i.e., p-values: 0.02 and 0.0005, respectively), whereas no statistical significance was observed compared to the Reflection group.

We further examined the screen recordings and GPT Chat logs to identify the distinguishing features of the Reflection and Few-Shot patterns in comparison to the other Alternative Approach and Cognitive Verifier patterns. We observed that the average number of requests for completing each class's development using the Reflection and Few-Shot patterns is lower than that of the Alternative Approach. For instance, in the Reflection and Few-Shot patterns, participants requested an average of 3.5 prompts and a maximum of 5 prompts to complete the code generation and fixing processes for a single class within the project task. In contrast, the other pattern groups had an average of 5.4 requests and a maximum of 7.3, requiring approximately 50\% more requests. This is primarily due to the internal processes of these patterns that necessitate participants responding to the sub-questions. While this difference may seem trivial in function-level tasks, in project tasks that involve iterative and progressive development and debugging of code within a limited timeframe, the inefficiency of requesting more prompts results in a relative decrease in productivity.

\begin{tcolorbox} The Few-Shot pattern demonstrated high productivity on average in both function and project-level SE tasks. The Reflection pattern yielded the best productivity in project-level tasks and received the highest subjective evaluations from participants. While the Cognitive Verifier pattern showed high productivity in function-level tasks, the Cognitive and Alternative Approach patterns are inefficient in iterative and progress SE tasks due to their interplay of execution processes.
\end{tcolorbox}

\subsubsection{RQ2-2: Quartile Group-Wise Analysis} \label{sec.expr.rq2-2}

In this RQ, we concentrated on thoroughly examining the prompting practices for the first quartile group in the project-level task experiment. During the function-level tasks, we observed that most participants interacted with GPT assistants approximately three times to complete a task; consequently, there were fewer than 20 communications across all function-level tasks, including initial implementation and fixing. In contrast, the project-level logs contain an average of 51.9 and a maximum of 82.3 requests, incorporating a variety of strategic prompting practices. We investigated the interaction features with GPT assistants in the first quartile group and compared them with the other participants to identify key differences.

Table~\ref{tab:RQ2} provides detailed information about the top six participants by analyzing their code outcomes, survey results, screen recordings, and chat logs with GPTs. The second through fourth columns display the experiment results for the top six participants and their assigned prompt patterns with GPT models. The fifth through seventh columns present background information about the participants. The eighth column highlights the types of context curation methods that the participants utilized alongside their predefined prompt patterns. The last three columns, Time distributions and Unique features, aim to explain the strategic aspects of prompting compared with the other quartile groups.

\begin{table*}
\caption{Experiment and survey results of the first-quartile group} \label{tab:RQ2}
\Large
\centering
 \resizebox{\textwidth}{!}{ 
\begin{tabular}{ccccccccccc}
\hline \hline
\Tstrut \multirow{2}{*}{Rank}     &  \multirow{2}{*}{TPR} & \multirow{2}{*}{\begin{tabular}[c]{@{}c@{}}Prompt \\ pattern\end{tabular}} & \multirow{2}{*}{GPT Tier} & \multirow{2}{*}{\begin{tabular}[c]{@{}c@{}}Programming \\ experience\end{tabular}} & \multirow{2}{*}{\begin{tabular}[c]{@{}c@{}}Professional \\ experience\end{tabular}} & \multirow{2}{*}{\begin{tabular}[c]{@{}c@{}}algorithm-solving \\ experience\end{tabular}} & \multirow{2}{*}{\begin{tabular}[c]{@{}c@{}}Context \\ curation\end{tabular}} & \multicolumn{2}{c}{Time Distribution} & \multirow{2}{*}{\begin{tabular}[c]{@{}c@{}}Unique \\ Strategy\end{tabular}} \\ \Bstrut
     &                                   &                                                                         &                      &                                                                                                                            &                                                                                     &                                                                                          &                                                                & Development     & Debugging     &                                                                      \\ \hline
1 & 0.533                           & Reflection                                                      & GPT-Paid                                  & 4-5 years                                                                          & 1-2 years                                                                          & \begin{tabular}[c]{@{}c@{}} \Tstrut \textless 5 \\ \Bstrut problems\end{tabular}                          & \begin{tabular}[c]{@{}c@{}}\Tstrut CopyPaste,\\ \Bstrut Formulate\end{tabular} & 19.17               & 50.83              & \begin{tabular}[c]{@{}c@{}}\Tstrut Test Case \\ \Bstrut as Context\end{tabular}      \\ \hline
2 & 0.520                           & Reflection                                                      & GPT-Paid                      & 6-8 years                                                                          & 3-4 years                                                                             & \begin{tabular}[c]{@{}c@{}}\Tstrut 6 to 10 \\ \Bstrut problems\end{tabular}                              & \begin{tabular}[c]{@{}c@{}}\Tstrut CopyPaste,\\ \Bstrut Formulate\end{tabular} & 15                  & 55              & \begin{tabular}[c]{@{}c@{}}\Tstrut Test Case \\ \Bstrut as Context\end{tabular}      \\ \hline
3 & 0.520                           & Few-Shot                                                        & GPT-Paid  & 4-5 years                                                                          & 1-2 years      & \begin{tabular}[c]{@{}c@{}} \Tstrut 20+ \\ \Bstrut problems\end{tabular}                                  & \begin{tabular}[c]{@{}c@{}} \Tstrut CopyPaste,\\ \Bstrut Formulate\end{tabular} & 27                  & 43              & \begin{tabular}[c]{@{}c@{}} \Tstrut Make GPT \\ \Bstrut Understand First\end{tabular} \\ \hline
4 & 0.515                           & Few-Shot                                                        & GPT-Paid    & 2-3 years                                                                          & None         & \begin{tabular}[c]{@{}c@{}} \Tstrut 11 to 15 \\ \Bstrut problems\end{tabular}                             & \begin{tabular}[c]{@{}c@{}} \Tstrut GPT-driven,\\ \Bstrut Formulate\end{tabular}                                                     & 13.5                  & 56.5            & \begin{tabular}[c]{@{}c@{}} \Tstrut Test Case \\ \Bstrut as Context\end{tabular}      \\ \hline
5 &  0.480                           & Few-Shot                                                        & GPT-Paid & 6-8 years                                                                          & 3-4 years                                                                                 & \begin{tabular}[c]{@{}c@{}} \Tstrut 6 to 10 \\ \Bstrut problems\end{tabular}                              & CopyPaste                                                      & 15                  & 55              & \begin{tabular}[c]{@{}c@{}} \Tstrut Test Case \\ \Bstrut as Context\end{tabular}      \\ \hline
6 & 0.480                           & \begin{tabular}[c]{@{}c@{}} \Tstrut Alternative \\ \Bstrut Approach\end{tabular} & GPT-Paid  & 2-3 years                                                                          & 1-2 years                    & \begin{tabular}[c]{@{}c@{}} \Tstrut \textless 5 \\ \Bstrut problems\end{tabular}                          & CopyPaste                                                      & 29.17               & 40.83              & \begin{tabular}[c]{@{}c@{}} \Tstrut Test Case \\ \Bstrut as Context\end{tabular}  \\   
\hline \hline
\end{tabular}
}
\end{table*}

\textbf{Context Curation Methods.} The first aspect of prompting strategies, alongside the prompt pattern, is the context curation method, which determines how participants crafted the prompt context. We categorized the primary curation methods into four types: CopyPaste, Manual Formulation, GPT-driven automated prompting, and other prompting patterns. As indicated in the eighth column of Table~\ref{tab:RQ2}, five out of the top six participants employed CopyPaste to extract contextual information from existing materials. Notably, the top three participants used CopyPaste for the context and integrated their understanding of the materials or specific requirements. 

For instance, the Rank-1 participant instructed the GPT to understand the prompt pattern template and strictly adhere to the Reflection pattern before beginning the task-solving. The participant outlined the expected behaviors and outcomes for the Reflection role to the GPT and then utilized our templates and materials to design task-solving prompts. This method is also evident in the logs of the Rank-3 participant, who set aside time in the beginning to explain the requirements and ensure that the GPT understood them before requesting code development or bug fixing, aiming to obtain more accurate answers. 
Besides, the Rank-4 participant employed GPT-driven automated prompt generation. The GPT was asked to design prompts after the participant provided the prompt template and all the design document data. The participant then directly used or manually crafted parts of the prompts generated by the GPT to generate and fix the code.

Based on the implications, we compared the TPR results from participants who only used CopyPaste with those who combined CopyPaste with the manual formulation. We found that the eight participants using CopyPaste and Formulating resolved an average of 95.88 test cases, with a minimum of 63 and a maximum of 121 test cases passed. In contrast, the other eight participants who solely copy-pasted the provided materials achieved an average of 77.25 test cases, with a minimum of 36 and a maximum of 109 test cases. Although we could not find statistically significant differences in their distributions from the Wilcoxon Ranksum test~\cite{wilcoxon1992individual}, this still suggests that enhancing the understanding of the provided contexts or specifying clear roles and expected outcomes can increase the code generation productivity by approximately 24\% on average. 

\begin{tcolorbox} 
While the accessible CopyPaste method for adding reference context using provided materials has been widely used, our experimental results explain that manually formulating additional context can further enhance productivity when interacting with GPT assistants. The manual method is particularly effective for clarifying your understanding of the requirements and issues, or specifying concrete behaviors and expected outcomes.
\end{tcolorbox}

Next, we conducted several statistical evaluations as part of our detailed investigation. These included the Wilcoxon Ranksum test, Fisher's exact test~\cite{fisher1922interpretation}, Cramer's V test~\cite{cramer1999mathematical}, and Cliff's $\delta$ test~\cite{cliff1993dominance}. We compared the HLI features of the first-quartile group with those of the other participants, including all user-side, model-side, and interaction-side features. The results revealed that the \textbf{GPT Tier} and \textbf{Time Distribution} features showed statistically significant differences between the data distributions of the first-quartile group and those of the other participant groups. Based on these statistical findings, we also identified differences in context artifact types among participants in the first-quartile group through a careful examination of their screen recordings and chat logs.

\textbf{GPT Tier.} We found that all participants in the first quartile group used the Paid 4o model, as outlined in Table~\ref{tab:RQ2}. Additionally, we observed significantly different distributions of GPT models (i.e., p-value: 0.001, Cramer's V: 0.539 — Large, post-hoc power: 85\%) between the first-quartile and other groups from the statistical analysis.
This result might raise a concern that the GPT Tier is the most significant factor in our experiment. However, as described in Section~\ref{sec.expr.rq1}, the GPT Tier feature does not present statistical significance or confidence in its impact on TPR values compared with prompt patterns and strategic time distribution features. Additionally, the other 12 Paid model users are distributed across the second-, third-, and fourth-quartile groups, while the second-quartile group contains four of the six Free model users. The details of the experimental analysis on the GPT Tier feature are further discussed in Section~\ref{sec.disc}.

\begin{tcolorbox} 
While Paid GPTs may offer better productivity opportunities, this has not been shown to significantly improve code generation productivity.
\end{tcolorbox}

\begin{figure}[t]
    \centering
    \includegraphics[width=0.65\textwidth, trim = 0.5cm 0.2cm 0.5cm 0.3cm,clip]{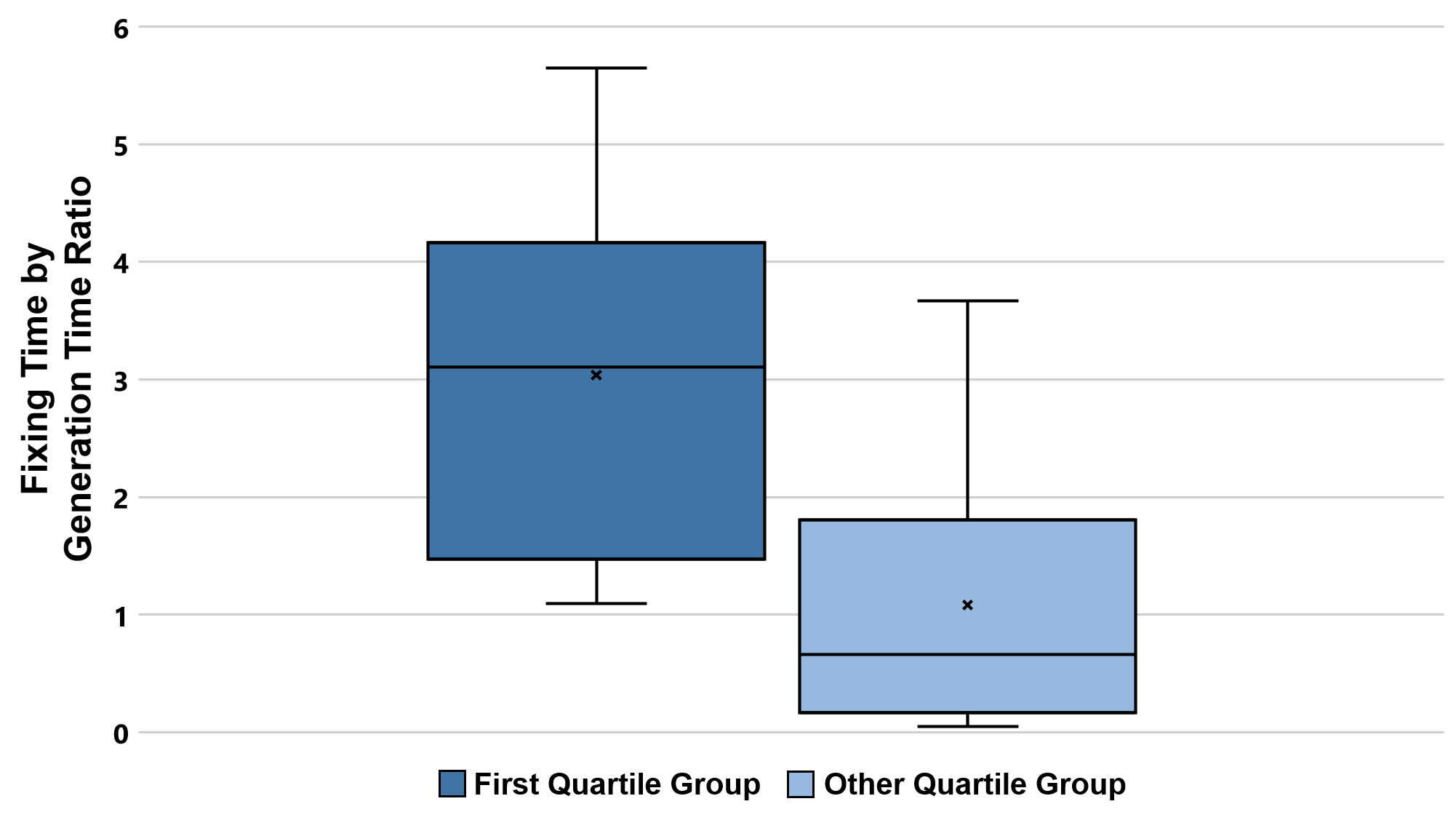}
    \caption{Initial implementation and code fixing time ratio of different quartile groups} \label{fig:RQ2}
    \vspace{-10px}
\end{figure}

\textbf{Time Distribution in Initial Development and Debugging Phases.} We measured participants' time distributions on initial code generation and iterative code fixing for the generated codes. Our statistical analysis results regarding the ratio of time spent on fixing versus generating code revealed that the first quartile group dedicated significantly more time to debugging the generated code than the other 18 participants (i.e., Wilcoxon Ranksum p-value: 0.014, Cliff's $\delta$: 0.73 — Very Large, post-hoc power: 82\%). Fig.~\ref{fig:RQ2} shows the ratio of minutes participants spent on code fixing compared to initial code generation. On average, the first quartile group spent three times as many minutes in the fixing phase as in the initial generation phase. This indicates they utilized approximately 17 minutes for code generation based on the design document and then allocated 50 minutes to refine the code based on test results. In contrast, most participants in the other quartile group spent about half the time generating their initial code, which still contained several errors, and they could not resolve them within the given time.

\begin{tcolorbox} 
It is advisable to swiftly transition to the code-fixing phase and iteratively refine the code with GPT assistants. GPT-generated codes, especially those involving multiple classes, often contain several faults, even after users spend significant time generating them.
\end{tcolorbox}

\textbf{Context Artifact.} Another feature we observed during the debugging phase is the context artifact that participants provided GPTs to explain failures. We found that participants primarily used design documents or test case codes. As noted in the last column of Table~\ref{tab:RQ2}, most participants in the first quartile used test results and test case codes to debug the failures. For example, the Rank-5 participant asked the GPT to filter the list of failed test case messages and address them individually by copy-pasting each message along with the corresponding test case code and specific class codes. Furthermore, we found that 10 participants in the top 12 ranks primarily used the test case codes instead of the documentation during the fixing phase.

\begin{tcolorbox} 
Our analysis revealed that utilizing test case codes can facilitate more accurate debugging of the code with GPTs than relying on design documents.
\end{tcolorbox}

\subsection{RQ3: Empirical HLI Error Analysis} \label{sec.expr.rq3}

The last RQ focuses on examining the identified errors in our experiment. In our user study, participants executed a total of 6,336 test cases. They encountered 791 errors, consisting of 420 runtime errors, 255 logic errors (i.e., failures), and 116 non-implementation errors. Except for the non-implementation errors, we performed an empirical analysis of 675 runtime and logic errors. For each error instance, three of the four authors independently reviewed the code outcome using the relevant GPT chat log and screen recordings. Each reviewer first assigned a symptom label and a candidate root cause; disagreements were then resolved through adjudication meetings until the team reached a consensus label. The reported taxonomy and frequencies, therefore, reflect the triage consensus.

Based on our analysis results, we established a taxonomy comprising 19 categories for runtime errors and 10 categories for logic errors that can occur during HLI processes in multi-class code generation tasks. Each category is defined by a combination of its symptoms, such as TypeError, LogicError, and its root causes. Furthermore, we identified 25 independent root causes, excluding four overlapping root causes that were found in both the runtime and logic error categories within HLI processes. 

Table~\ref{tab:RQ3_1} and \ref{tab:RQ3_2} describe the taxonomy of runtime and logic errors analyzed in this study, respectively. The Name column presents the plain-language description of identified errors, the Frequency indicates the number of detected errors in our experiment, and the last column explains how the root causes induce the particular errors. We explain these errors and their root causes based on the origins of faults from Users, Models, and Models during Debugging.

\begin{table*}[t]
\centering
\Large
\caption{Taxonomy of Runtime Errors Identified in Our Experiment} \label{tab:RQ3_1}
\resizebox{0.98\textwidth}{!}{
\begin{tabular}{llll}
\hline \hline
\TBstrut Runtime Error                   & Error Name                                                                                                    & Frequency & Explanation                                                                                                                                                            \\ \hline
\multirow{3}{*}{}      & [T-1] Inconsistent Data Structure Handling                                                                    & 16        & \begin{tabular}[c]{@{}l@{}} \Tstrut The initialization and usage of data structures are inconsistent \\ \Bstrut (e.g., Initiating List, Treating Dictionary)\end{tabular}             \\
                          TypeError      & [T-2] Wrong Output Format                                                                                     & 7         & \begin{tabular}[c]{@{}l@{}} \Tstrut Returned output format is hallucinated due to the lack of context \\ \Bstrut (List expected, but Dictionary developed)\end{tabular}                 \\
                                & [T-3] No dependency with other classes                                                                        & 16        & \begin{tabular}[c]{@{}l@{}} \Tstrut A class contains List or Dictionary object named with\\ \Bstrut dependent Classes, not the Class objects \end{tabular}                             \\ \hline
\multirow{6}{*}{} & [A-1] Missing Attribute                                                                                       & 22        & \TBstrut Attributes required in certain Classes are omitted.                                                                                                                    \\
                                & [A-2] Wrong Parameter Type in Function Call                                                                   & 24        & \TBstrut Different types of parameters are used in calling dependent functions                                                                                                  \\
                         & [A-3] Immutable Attribute Setting                                                                             & 70        & \begin{tabular}[c]{@{}l@{}} \Tstrut Attributes have been defined as Immutable, \\  \Bstrut  while they should be Mutable.\end{tabular}                                                  \\
                    AttributeError & [A-4] Existing Attribute Overlapped                                                                           & 22        & \TBstrut Some of the existing Attributes are removed in fixing specific Classes                                                                                                     \\
                                & [A-5] Parameter Order Mixed                                                                                   & 19        & \begin{tabular}[c]{@{}l@{}} \Tstrut In fixing a specific function, the parameter order of\\ \Bstrut the function is unexpectedly changed.\end{tabular}                                \\
                                & [A-6] Condition Checking Order Issue                                                                          & 4         & \begin{tabular}[c]{@{}l@{}} \Tstrut String processing is executed before checking \\ \Bstrut whether the input type is String\end{tabular}                                            \\ \hline
\multirow{7}{*}{}     & [V-1] Wrong Regex Condition                                                                                   & 66        & \begin{tabular}[c]{@{}l@{}}\Tstrut Case sensitivity, Strip, and other conditions of Regex \\  \Bstrut on Email address are not correctly developed.\end{tabular}                      \\
                                & [V-2] Wrong AND, OR Condition                                                                                 & 30        & \begin{tabular}[c]{@{}l@{}} \Tstrut String parsing and checking should be AND conditions, \\ \Bstrut but the strip() is ignored or checked with OR conditions. Vice versa.\end{tabular} \\
                                & \begin{tabular}[c]{@{}l@{}}[V-3] \Tstrut Hallucinated Conditions and \\     \Bstrut      Unexpected Error Raised\end{tabular} & 26        & \begin{tabular}[c]{@{}l@{}} \Tstrut Insufficient context given to GPT, GPT generates the functions\\ \Bstrut  with hallucinated conditions.\end{tabular}                               \\
                    ValueError            & [V-4] Hallucinated Attributes Generated                                                                       & 1         & \begin{tabular}[c]{@{}l@{}} \Tstrut GPT generates hallucinated attributes that are not requested \\ \Bstrut to include by Users.\end{tabular}                                         \\
                                & [V-5] Wrong Boundary Setting                                                                                  & 45        & \begin{tabular}[c]{@{}l@{}} \Tstrut Wrong boundary conditions for float, integer variable have \\ \Bstrut been implemented.\end{tabular}                                              \\
                                & [V-6] Missing Conditions                                                                                      & 17        & \TBstrut Required conditions are omitted in the implementation.                                                                                                                 \\
                                & [V-7] Wrong Return Type                                                                                       & 14        & \begin{tabular}[c]{@{}l@{}} \Tstrut Expected to return specific data structure (e.g., List), \\ \Bstrut but actually return object in different format. (e.g., Dictionary)\end{tabular} \\ \hline
\multirow{3}{*}{}       & [K-1] Wrong Key Type                                                                                          & 18        & \begin{tabular}[c]{@{}l@{}} \Tstrut Dictionary key type is developed wrongly. \\ \Bstrut (e.g., Expected: Numeric ID key, Code: String Name key)\end{tabular}                           \\
                    KeyError            & [K-2] Wrong Key ID Starting Number                                                                            & 3         & \begin{tabular}[c]{@{}l@{}} \Tstrut 0-based ID numbering is expected, but IDs starting \\ \Bstrut from 1 have been applied.\end{tabular}                                              \\
                                & [K-3] Wrong Return Type                                                                                       & 10        & \begin{tabular}[c]{@{}l@{}} \Tstrut Return Boolean is expected, but Return Integer, \\ \Bstrut especially zero that is considered False, has been implemented.\end{tabular}   \\          
\hline \hline
\end{tabular}
}
\vspace{-15px}
\end{table*}

\begin{table*}[t]
\Large
\centering
\caption{Taxonomy of Logic Errors (i.e., Failures) Identified in Our Experiment} \label{tab:RQ3_2}
\resizebox{0.98\textwidth}{!}{
\begin{tabular}{llll}
\hline \hline
 \multirow{11}{*}{} \TBstrut & Error Name                                                                                                         & Frequency & Explanation                                                                                                                                                                                                     \\ \cmidrule{2-4}
                                              & [L-1] Wrong Return Type                                                                                            & 10        & \begin{tabular}[c]{@{}l@{}} \Tstrut Expected to return Boolean, but return Integer, \\ \Bstrut  especially 0 that is considered False.\end{tabular}                                                                               \\
                                              & [L-2] Rounding Issue                                                                                               & 4         & \begin{tabular}[c]{@{}l@{}} \Tstrut Too strict rounding strategy on floating values,\\ \Bstrut which is not required.\end{tabular}                                                                                               \\
                                              & [L-3] String Cleaning before Validation                                                                            & 2         & \begin{tabular}[c]{@{}l@{}} \Tstrut String values are preprocessed and \\ \Bstrut cleaned before the validation.\end{tabular}                                                                                                    \\
                                              & [L-4] Existing Condition Overlapped                                                                                & 12        & \begin{tabular}[c]{@{}l@{}} \Tstrut Existing conditions are omitted by refining codes \\ \Bstrut for satisfying different test cases\end{tabular}                                                                                \\  Logic Error (i.e., Failure)
                                              & [L-5] Existing Function Name Overlapped                                                                            & 14        & \begin{tabular}[c]{@{}l@{}} \Tstrut Existing function names are overlapped by random \\ \Bstrut function names during debugging phase\end{tabular}                                                                               \\
                                              & [L-6] Expected to Raise Error, but No Error Raised                                                                 & 58        & \TBstrut Wrong conditions failed to handle exceptional inputs                                                                                                                                                            \\
                                              & [L-7] Missing Conditions                                                                                           & 61        & \begin{tabular}[c]{@{}l@{}} \Tstrut Conditions for checking ranges of numbers or string \\ \Bstrut exceptions are omitted in implementation\end{tabular}                                                                         \\
                                              & [L-8] No dependency with other classes                                                                             & 66        & \begin{tabular}[c]{@{}l@{}} \Tstrut One class contains List/Dictionary named with \\ \Bstrut dependent Classes, not the Class objects\end{tabular}                                                                               \\
                                              & \begin{tabular}[c]{@{}l@{}}[L-9] Function Location Error \\             in Debugging Multiple Classes\end{tabular} & 15        & \begin{tabular}[c]{@{}l@{}} \Tstrut Move one function from the originally located class \\ to another class, and Generate a new function that works \\ \Bstrut similarly with different names in the original class\end{tabular} \\
                                              & [L-10] Wrong Boundary Setting                                                                                      & 13        & \TBstrut Wrong boundary for integer variable was set     
 \\          
\hline \hline
\end{tabular}
}
\vspace{-15px}
\end{table*}

\textbf{User-Originated Errors.} There are 11 categories of runtime and logic errors that primarily arise from the user's provision of insufficient context, mistakes, or misunderstandings regarding the content of the design document.
For instance, the [T-1] errors in Table~\ref{tab:RQ3_1} occur when participants copied and pasted only certain parts of the code, specifically omitting the initialization of new objects, from the responses generated by the GPT in the WebUI. This led to conflicts between the original initialization of list objects and the usage of new objects in the final code. 
Other error types, such as [A-3] and [L-10], stem from users' misunderstandings of the design document as follows: 
In the Cognitive Verifier pattern, GPTs inquired about the “Immutability” of attributes within specific classes, but participants provided incorrect responses. Additionally, in the case of [L-10] errors, one participant mistakenly defined the acceptable string conditions as “1 or 50” instead of the correct range of “1 to 50.”

While we found that these minor user-side faults result in fatal errors, most user-originated errors are caused by GPTs' hallucinations when faced with incomplete context while developing functions and classes.  There are eight types of errors associated with this issue, categorized as [T-2,3], [A-5], [V-3,5], and [L-1,6,7], all of which arise from insufficient information provided by the users. 
For instance, [T-2] errors occur when participants fail to specify the output data structure required for specific functions. The hallucination due to a lack of context also occurs when unclear descriptions of conditions within various functions are given. Another example is the multi-class dependency error [T-3], which occurred when participants did not include the definitions or code for dependent classes, such as \texttt{Product} or \texttt{ShoppingCart}, in their prompts when requesting the development of another class, like \texttt{EcommerceApp}.
However, we found that GPTs have already developed dependent classes within the same chat session. This indicates that although GPTs have generated classes with the same names previously, they have weaknesses in handling multi-class dependencies intelligently, resulting in the implementation of List or Dictionary attributes with the same names as previously developed classes.

In our analysis of insufficient context errors, we identified another pattern: participants with a higher score of algorithm-solving Experience (defined as a score greater than 75) typically created context manually during the initial development phase, rather than copying and pasting directly from the design description. Although some participants copied and pasted context from the provided document during the debugging phase, the GPTs struggled to identify the appropriate fixes for code that already contained hallucinated structures. This indicates that relying solely on manual context formulation may lead to hallucinations due to insufficient information about the functions and classes involved.

\begin{tcolorbox} 
While minor mistakes and misunderstandings of context from users can lead to a significant number of errors, it is advisable for GPT users to provide sufficient context when developing target functions or classes, especially for multi-class dependencies. Manual context formulation is not recommended without copying and pasting the original specification, as it increases the risk of hallucinations on missing details.
\end{tcolorbox}

\textbf{Model-Originated Errors.} The other 18 categories of errors arise from the misbehavior of GPTs, despite users providing the appropriate context in their requests. Among these 18 types, 13 error patterns occurred during the initial implementation phase, as listed in Table~\ref{tab:RQ3_1} as [A-1,2], [V-1,2,4,6], and [K-1,2], and in Table~\ref{tab:RQ3_2} as [L-2,3,8]. 

AttributeErrors: [A-1,2] primarily result from missing required attributes or the incorrect use of parameter types in specific functions when they are called. For example, the \texttt{\_init\_()} method in the \texttt{Order} class is defined to accept a parameter \texttt{ItemList: List}, but GPT-generated code incorrectly calls the method with a parameter of another class, \texttt{ShoppingCart}, instead of the expected List.
Another attribute-related cause of ValueError, noted as [V-4], appears in the following scenario: A participant requests the development of a Product class with adequate context, but GPT unexpectedly inquires about stock management, which was not included in the request. When the participant clarifies that stock management should not be considered, GPT still generates this attribute in the class and introduces stock-related conditions, resulting in value errors.

Most of the root causes of the ValueError stem from issues in developing incorrect conditions from GPTs. Specifically, 66 ValueErrors are caused by the incorrect regex condition settings in [V-1], while 30 ValueErrors arise from improper combinations of AND and OR in a single if statement in [V-2]. Additionally, cases in [V-6] reveal missing conditions from the required sets. Many of these errors are related to string processing functions, such as \texttt{String.strip()} and \texttt{String.lower()}. These functions are often implemented in an OR relationship within a conditional statement when they should be combined with an AND. 

An interesting root cause of the missing condition errors arises from the use of the \textit{Alternative Approach} pattern. We found that when GPTs generate three different answers based on the prompt pattern, the in-depth context information given by the users is only applied to the first answer. In contrast, the two subsequent answers often contain hallucinated conditions appearing to be developed without the concrete conditions. However, in all these cases, GPTs tend to recommend the second or third answer, which contains these hallucinated conditions, highlighting the persuasive strengths of its structure. Participants in our experiment generally followed these suggestions.

Similarly, KeyErrors are mainly caused by (1) improper implementation of the Key type in a Dictionary about the specified requirements, and (2) counting IDs starting from 1 instead of 0.

In LogicErrors, we found that these error types of [L-2,3] are caused by similar root causes to other RuntimeErrors, except for the [L-8] error. For example, in the first project-level task, this error returns the main class, \texttt{EcommerceApp}, with no or partial connection to other preceding classes, such as \texttt{Product} and \texttt{Order}. Instead, GPTs generated \texttt{List} or \texttt{Dictionary} objects with the same name and save corresponding information in those ad-hoc objects. The key distinction between these errors and those categorized as User-originated errors is that, in this case, relevant information about the classes was provided in the request. Nevertheless, GPTs were unable to develop the multi-class dependency correctly. Our empirical inference suggests that the primary cause of these errors was related to the context window size. Errors occurred when participants copied and pasted the entire content of the design document while asking for the development of the \texttt{EcommerceApp}. Because the \texttt{EcommerceApp} uses all four other classes, the coverage of information used by the participants may not be incorrect. Still, they appended too many redundant details of the comprising functions in each class. These findings further clarify the superior performance of the \textit{Few-Shot} pattern, which includes well-abstracted information for one of the dependent classes in its example content. 

\begin{tcolorbox} 
Even with the proper context, GPTs still have the uncertainty of producing erroneous code. This uncertainty increases when users request multiple solutions for optimized decision-making or when they get excessive and redundant details. It is essential to clearly describe the necessary context and allocate sufficient time for thorough debugging processes to minimize uncertainties.
\end{tcolorbox}

\textbf{Errors Occurred in Debugging Phase.} The remaining five categories of errors are described in Table~\ref{tab:RQ3_1} as [A-4,5] and in Table~\ref{tab:RQ3_2} as [L-4,5,9]. These errors are all caused by the GPTs in generating patches of existing code. For example, the [A-4,5] errors arose from the unexpected removal of attributes in fixing and the change of parameter orders in specific function calls, respectively. Similarly, [L-4,5] errors have root causes of (1) omission of certain conditions in corresponding functions and (2) name changes of functions. These errors share a common point in that the newly added error parts are not directly relevant to the actual fixing locations.

The [L-9] error presents a multi-class dependency issue in fixing logic related to multiple classes. In this error, the \texttt{add\_to\_cart()} function, located initially in the \texttt{ShoppingCart} class, is moved to another \texttt{User} class, followed by adding fixed logic in a new function called \texttt{add\_items()} in the \texttt{ShoppingCart} class. This indicates that the uncertainty in GPTs can lead to confusion about the multi-class dependencies and induce side-effect errors.

\begin{tcolorbox} 
During the debugging phase, GPT users should be aware of the risk of code overlap with the original code, particularly in multi-class fixing scenarios, where the side-effect errors need to be carefully reviewed. 
\end{tcolorbox}

\subsection{Threats to Validity} \label{sec.expr.TV}

\textbf{Internal Validity.} Some screen recordings or GPT chat logs are unavailable for certain participants. For instance, one participant who joined via their Ubuntu environment could not use the screen recording tool we prepared. Consequently, we concentrated on examining GPT chat logs for the three participants who could not provide a screen recording. However, the communication logs from OpenAI lack time stamps, so we inferred the approximate timeline for the code generation and fixing phases based on their prompt logs. Additionally, some issues exist in GPT logs due to OpenAI's logging limitations. When we have restricted access to request and answer logs, we primarily use the screen recording files. Additionally, to ensure serious engagement from participants in our experiments, we distributed the rewards: 50 USD for 90 minutes of participation after reviewing all the submitted files.

\textbf{External Validity.} 
The project-level tasks were intentionally designed as controlled dependency-rich projects rather than as full replicas of an industrial codebase. Likewise, all participants used the same web interface to avoid interface effects dominating the study. Accordingly, the findings would be most effective in conversational code-assistant use under controlled conditions and may transfer only partially to modern IDE-integrated assistants, large external APIs, collaborative team workflows, and codebases with substantial hidden context or organization-specific tooling. We therefore interpret the study as evidence about interaction behavior under controlled task conditions rather than as a complete model of real-world software development productivity.


\textbf{Construct Validity.} Our study defined productivity by considering the accuracy of outcomes, task-solving efficiency, and subjective evaluations of development satisfaction. However, traditional productivity measures also consider maintenance, security, and reusability~\cite{boehm1987improving,scacchi1995understanding,mccabe1976complexity,ISO25002:2024}. We anticipate that this future analysis will provide insights into which GPTs or prompting strategies are likely to produce high-complexity, vulnerable, or duplicated code.

Additionally, this study allowed for limited manual coding after three unsuccessful attempts to resolve the same issue using GPT requests. As a result, the final TPR reflects outcomes from a combination of GPT-generated code, iterative interactions between humans and GPT, and, in rare instances, participant corrections. Therefore, we view the TPR as a comprehensive measure of task performance within the controlled experiment, rather than as a standalone indicator of productivity generated by GPT.

Lastly, by the experiment design, access to Task 2 was conditional on passing 121 out of 128 Task 1 tests. The tasks are designed to cover the full capability of HLI-based implementations within 70 minutes. However, in this experiment, most participants could not meet the Task 1 gate thresholds; thus, they remained in partial completion of Task 1 and did not reach Task 2. We believe future experiments can address the combinatorial analysis of Tasks 1 and 2 outcomes.

\textbf{Conclusion Validity.} 
Our quantitative analysis is exploratory. With 36 participants and 15 candidate features, the study was not powered to perform precise confirmatory inference across all pairwise groupings. A sensitivity analysis of our outcome shows that balanced two-group comparisons of 18 vs 18 participants would mainly detect large effects (about $d=0.96$); pairwise prompt-pattern contrasts of 9 vs 9 participants would require very large effects (about $d=1.41$); and full-sample correlations are mainly detectable when they are moderate-to-large. We therefore interpret the statistical results as supportive, associative evidence and deliberately triangulate them with descriptive results, effect sizes, screen recordings, chat logs, and manual inspection of submitted codes.

In addition, during the analysis of error taxonomy, three authors independently reviewed each error through adjudication. However, we did not keep the original pre-adjudication label matrix in a format that would allow for a numerical agreement figure. The reported taxonomy reflects the consensus reached through adjudication among the authors and should be interpreted in light of this limitation.

\section{Discussion} \label{sec.disc} 

\begin{figure}[t]
    \centering
    \includegraphics[width=0.65\textwidth, trim = 0.5cm 0.5cm 0.7cm 0.4cm,clip]{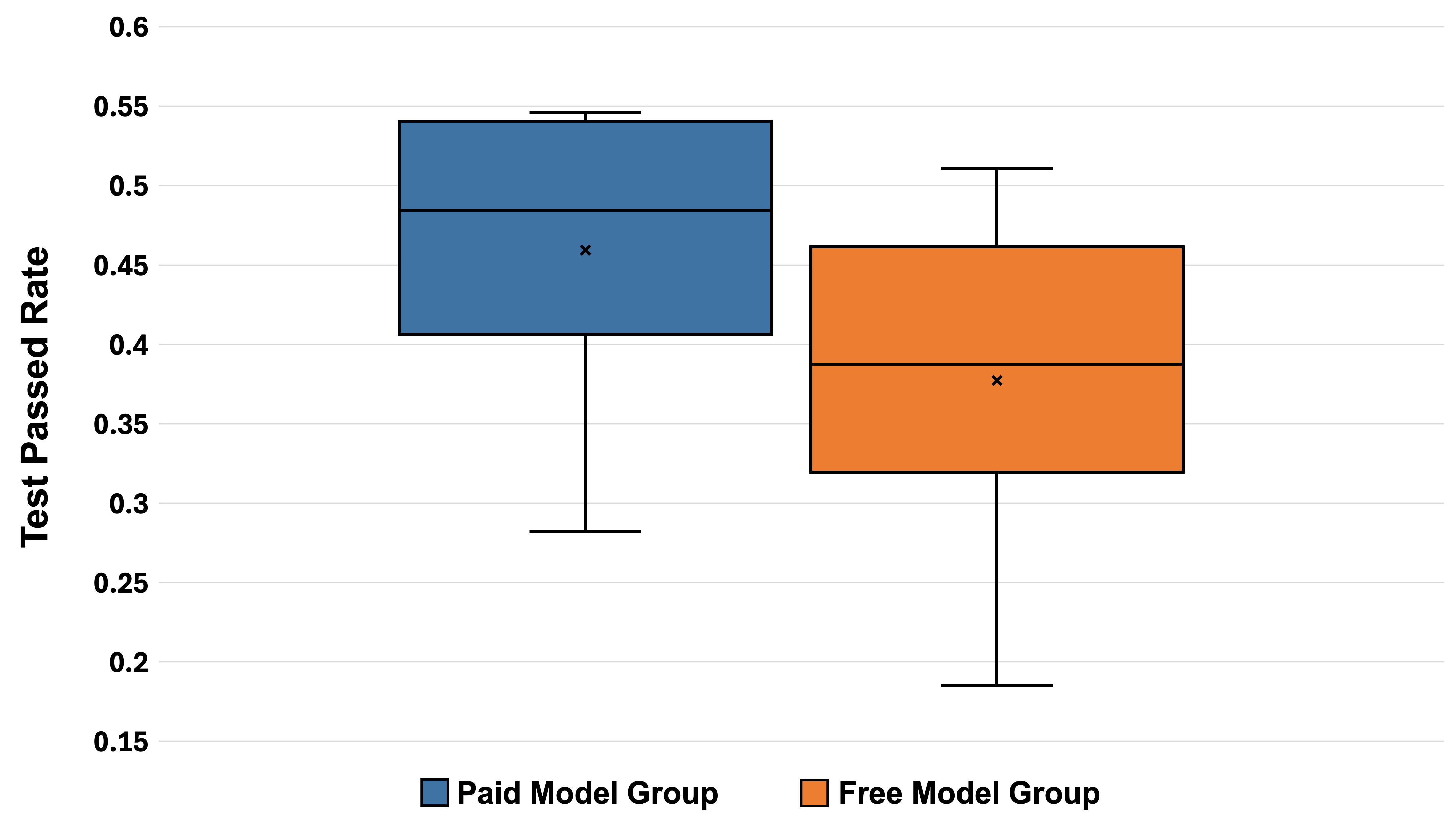}
    \caption{TPR Comparison with Paid and Free GPT Groups} \label{fig:discussion}
    \vspace{-15px}
\end{figure}

\textbf{General Discourse: Better Model vs Better Prompting.} In LLM applications, there is an ongoing debate about whether to focus on improving the models themselves or on developing more effective prompting strategies to enhance performance. The investigation of our experiment results revealed that while both aspects are synergistic, the prompting strategies should be emphasized more in the code generation task. 

Fig.~\ref{fig:discussion} illustrates the code generation productivity results, TPR values, for the Paid and Free GPT groups among 24 participants in the project-level task. On average, the Paid group achieved a productivity level that is 12.2\% higher than that of the Free group. However, statistical analysis using the Wilcoxon rank-sum test yielded a p-value greater than 0.05, indicating no significant difference in distribution between the two groups. In contrast, we observed that different prompting strategies generated substantial differences within both the Paid and Free GPT groups. For example, the \textit{Few-Shot} prompting pattern outperformed all other patterns within the Free GPT group, although it had a power value of only 0.56. In the Paid group, Spearman's rank correlation test~\cite{spearman1961proof} revealed that the time distribution feature significantly affects TPR, with a p-value of 0.004 and a power of 0.84. These results indicate that there is a statistically significant difference in productivity across features within the same GPT group.

Based on the investigation outcomes from the RQs and the Paid-versus-Free comparison, the study suggests that prompting strategy and interaction behavior remain important sources of variation in productivity even when participants use the same conversational interface. At the same time, because the experiment standardized the interface on a web-based chat setting, we do not claim that similar patterns will transfer unchanged to other IDE-based or API-driven workflows in other industry environments. Instead, the present findings provide strong evidence that the interaction strategy still matters under controlled conversational code generation conditions.

\textbf{HLI Features on Different Roles in SE Process.} In the SE process, several roles exist, including Architecture Designer, Software Engineer, and Tester~\cite{lin2025soen}. Our study aimed to examine the one-to-one interactions of human engineers with GPT assistants for code generation. Therefore, we defined a set of 15 features that are known to impact and are expected to affect the code generation productivity during the HLI processes.
We believe this research scope can be expanded to include multi-role and multi-participant experiments to assess the GPTs' effectiveness in a more scalable environment with multilateral aspects. For instance, after fine-tuning GPT agents for various roles in SE processes, one or more roles can be substituted with human participants with specific experience. In this case, a different HLI feature space, including experience on specific roles (e.g., architecture designer) or combined sets of multiple prompting patterns for different roles, can be defined and assessed throughout the experiment.  We believe our project-level task design and a set of HLI features for code generation productivity can serve as a foundation for designing the multi-role user study, covering different roles in the SE process.

\section{Conclusion} \label{sec.conclusion}
Our study presents a controlled empirical assessment of how developers interact with GPT assistants in function-level and dependency-rich project-level code generation tasks. The main contribution of the paper is not limited to a universal ranking of prompt techniques, but introduces an evidence-backed characterization of which interaction behaviors appeared most productive under these study conditions. By combining task outcomes with exploratory statistical screening and detailed trace analysis, we identify a small set of candidate features that consistently merit attention in enhancing the code generation productivity.

Our findings pave the way for exploring more productive interactions between engineers, particularly students and early-career practitioners, and GPTs in project-level code generation tasks. For software engineers, these results emphasize the value of explicitly integrating requirement details and spending focused effort on code fixing after initial implementation. Team managers may focus on streamlining development process workflows and improving team efficiency, especially in large-scale projects. Researchers can utilize our project-level benchmarks and the error taxonomy to refine or propose new prompting paradigms that address the uncertainty and dependency issues more accurately or efficiently.

We expect to conduct several future studies based on the results and artifacts from this experiment. First, we can engage in multi-role and multi-participant experiments to gain insights into collective best practices and knowledge-sharing by extending the suggested feature space in the HLI processes, such as combining multiple prompting patterns and models for different sub-tasks (e.g., code generation and explanation). Lastly, a domain-specific exploration of concrete case scenarios, including machine-learning codes or specific quality attributes (e.g., security or maintenance), may reveal how prompts should be structured for the cases.

\newpage
\section*{Appendix}

\begin{figure}[h]
     \centering
     \begin{subfigure}[b]{\textwidth}
        \centering
         \includegraphics[width=0.3\textwidth, height=5cm, trim = 0.5cm 0.4cm 0.7cm 0.3cm,clip]{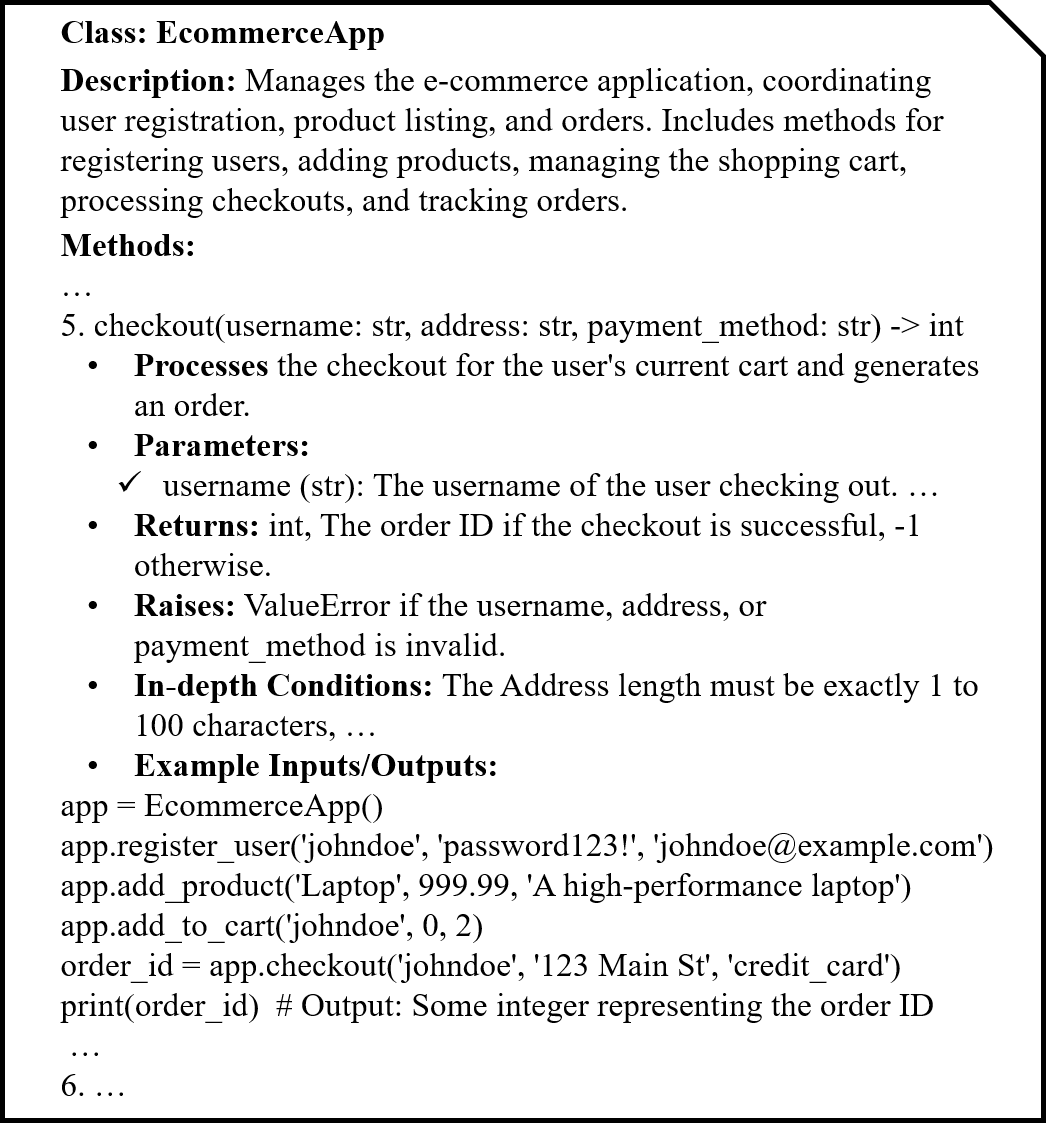}
         \caption{Example Design Description for \texttt{checkout()} function in \texttt{EcommerceApp} Class}
     \end{subfigure}
     \begin{subfigure}[b]{\textwidth}
        \centering
         \includegraphics[width=0.3\textwidth, height=6.1cm, trim = 0cm 0cm 0cm 0cm,clip]{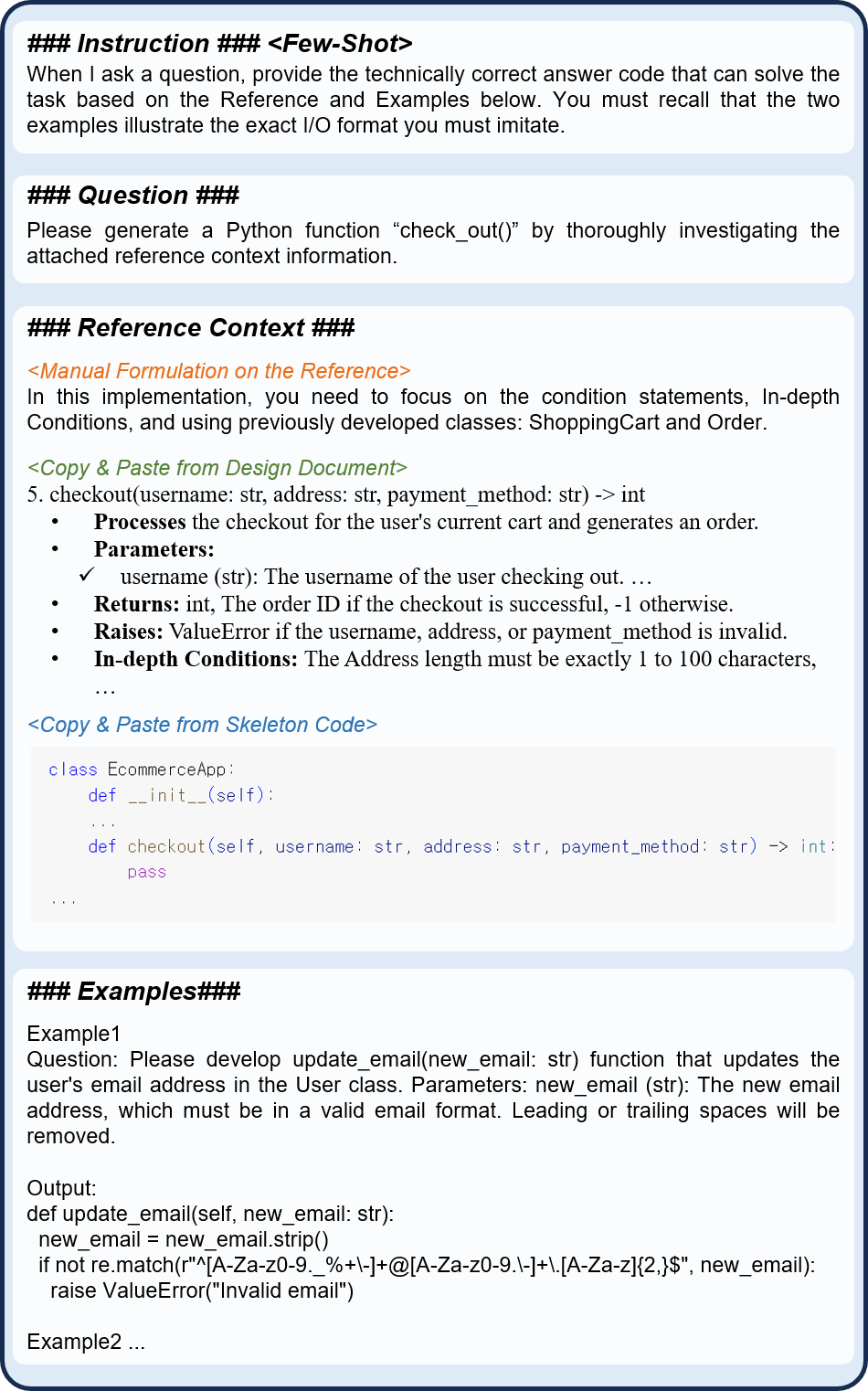}
         \caption{Example Few-Shot prompt pattern used by a participant in generating the \texttt{checkout()} function}
     \end{subfigure}
     \begin{subfigure}[b]{\textwidth}
    \centering
     \includegraphics[width=0.3\textwidth, height=5cm, trim = 0cm 0cm 0cm 0cm,clip]{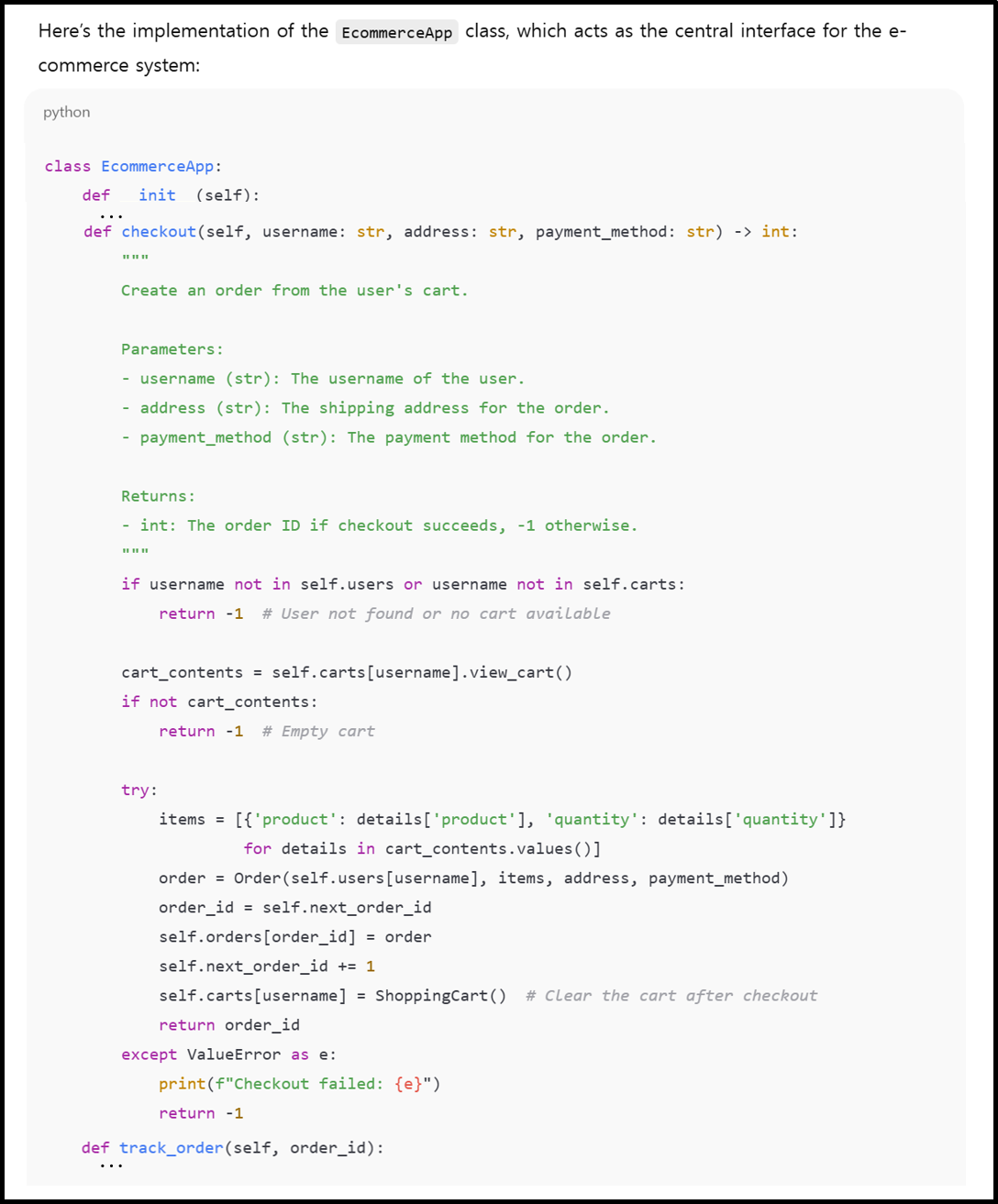}
     \caption{Example response of generating \texttt{checkout()} function}
     \end{subfigure}
     \caption{Example Code Generation Practice in Our Experiment}
\end{figure}


\newpage
\begin{figure}[h]
     \centering
     \begin{subfigure}[b]{\textwidth}
        \centering
         \includegraphics[width=0.45\textwidth, trim = 0.0cm 0.0cm 0.0cm 0.0cm,clip]{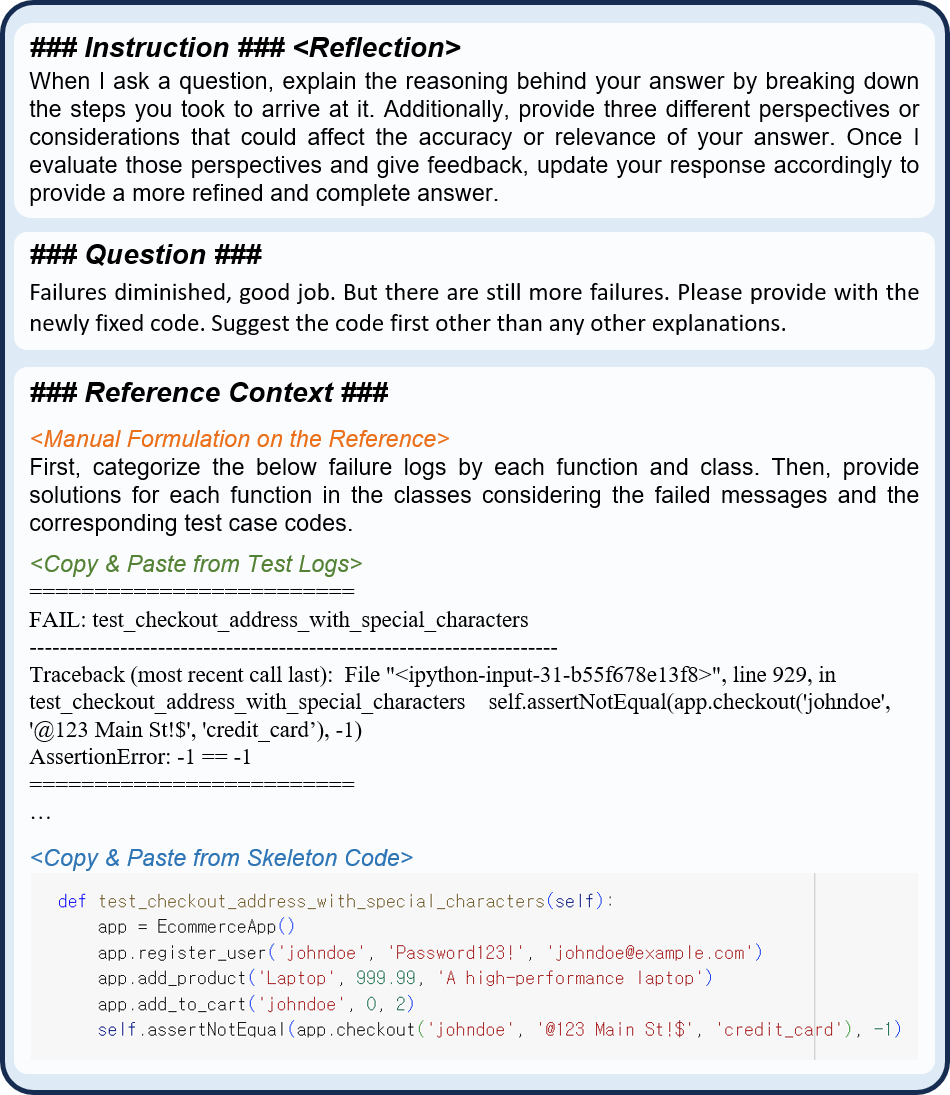}
         \caption{Example Reflection pattern used by a participant in fixing the \texttt{checkout()} function}
     \end{subfigure}
     \begin{subfigure}[b]{\textwidth}
        \centering
         \includegraphics[width=0.45\textwidth, height=5cm, trim = 0cm 0cm 0cm 0cm,clip]{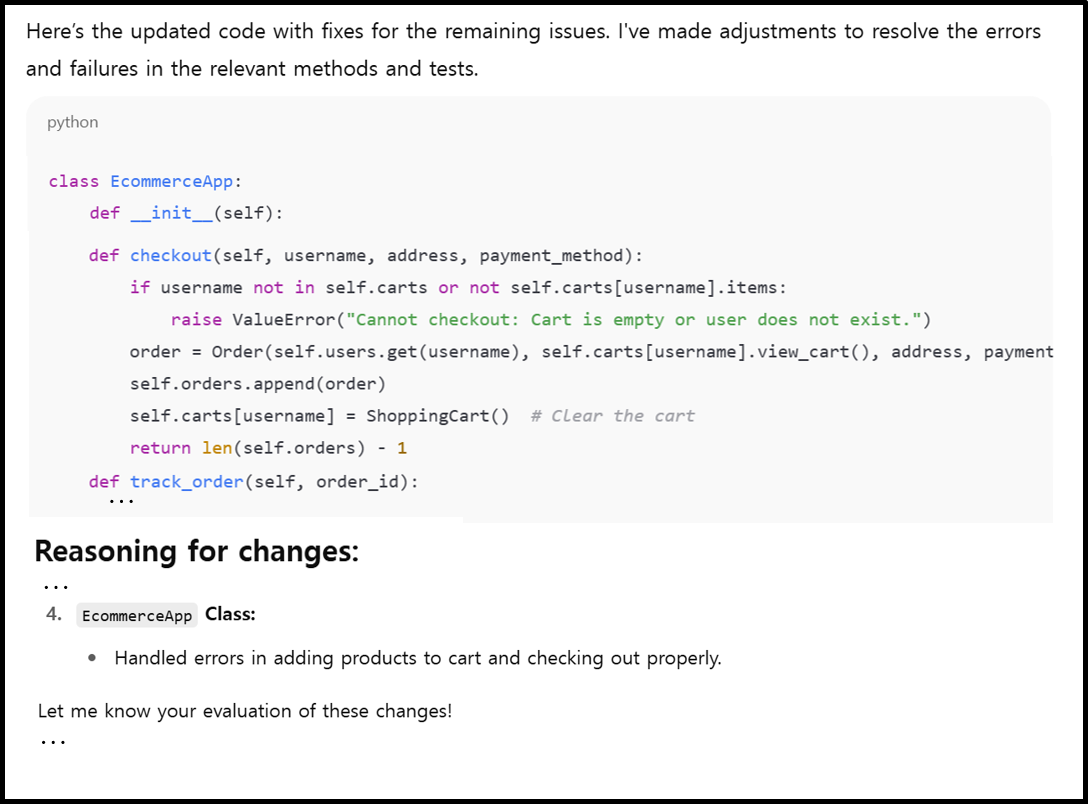}
         \caption{Example response of fixing the \texttt{checkout()} function}
     \end{subfigure}
     \caption{Example Debugging Practice in Our Experiment}
\end{figure}

\newpage
\section*{Acknowledgements}
We thank Alexander John Brown for his assistance in conducting the user studies and analyzing experiment results.

\section*{Declarations}
\textbf{Funding:} The work was supported by the Cyber Security Research Centre Limited, whose activities are partially funded by the Australian Government’s Cooperative Research Centres Program, and supported by the MSIT (Ministry of Science and ICT), Republic of Korea, under the National Program for Excellence in SW), supervised by the IITP (Institute of Information \& Communications Technology Planning\&Evaluation) in 2024 (2022-0-01092).
\newline
\\
\noindent
\textbf{Ethical Approval:} This study was supported by the Ethical Approval of H-2024-159 for Investigating Software Development Process by Using Large Language Models from the University of Adelaide.
\newline
\\
\noindent
\textbf{Informed Consent:} Not applicable.
\newline
\\
\noindent
\textbf{Author Contributions:} Sangwon Hyun - \textit{Conceptualization, Methodology, Software, Resources, Formal analysis and investigation, Writing - original draft preparation}, Hyunjun Kim, Jinhyuk Jang, Hyojin Choi - \textit{Methodology, Validation, Writing - review and editing}, M. Ali Babar - \textit{Writing - review and editing, Supervision, Funding acquisition, Project administration}
\newline
\\
\noindent 
\textbf{Data Availability Statement:} The datasets utilized and generated during this experiment, along with the benchmark information, can be found in our GitHub repository (https://github.com/abalon1210/Multi-Class-LLM-CodeGen-Benchmark). Please note that we cannot provide screen recording videos or GPT chat logs that contain personal information.
\newline
\\
\noindent
\textbf{Conflict of Interest:} 
All authors certify that they have no affiliations with or involvement in any organization or entity with any financial or non-financial interest in the subject matter or materials discussed in this manuscript.
\newline
\\
\noindent
\textbf{Clinical Trial Number:} Not applicable.

\bibliographystyle{ACM-Reference-Format}
\bibliography{sample-base}

@String{Computing = "Computing" }

@String{Computer = "{IEEE} Computer" }

@String{Springer = "Springer-Verlag" }

@ArtifactSoftware{R,
    title = {R: A Language and Environment for Statistical Computing},
    author = {{R Core Team}},
    organization = {R Foundation for Statistical Computing},
    address = {Vienna, Austria},
    year = {2019},
    url = {https://www.R-project.org/},
}

@article{wei2022chain,
  title={Chain-of-thought prompting elicits reasoning in large language models},
  author={Wei, Jason and Wang, Xuezhi and Schuurmans, Dale and Bosma, Maarten and Xia, Fei and Chi, Ed and Le, Quoc V and Zhou, Denny and others},
  journal={Advances in neural information processing systems (NeurIPS)},
  volume={35},
  pages={24824--24837},
  year={2022}
}

@inproceedings{gao2024taxonomy,
  title={A Taxonomy for Human-LLM Interaction Modes: An Initial Exploration},
  author={Gao, Jie and Gebreegziabher, Simret Araya and Choo, Kenny Tsu Wei and Li, Toby Jia-Jun and Perrault, Simon Tangi and Malone, Thomas W},
  booktitle={Extended Abstracts of the CHI Conference on Human Factors in Computing Systems},
  pages={1--11},
  year={2024}
}

@article{chen2021evaluating,
  title={Evaluating large language models trained on code},
  author={Chen, Mark and Tworek, Jerry and Jun, Heewoo and Yuan, Qiming and Pinto, Henrique Ponde De Oliveira and Kaplan, Jared and Edwards, Harri and Burda, Yuri and Joseph, Nicholas and Brockman, Greg and others},
  journal={arXiv preprint arXiv:2107.03374},
  year={2021}
}

@inproceedings{du2024evaluating,
  title={Evaluating large language models in class-level code generation},
  author={Du, Xueying and Liu, Mingwei and Wang, Kaixin and Wang, Hanlin and Liu, Junwei and Chen, Yixuan and Feng, Jiayi and Sha, Chaofeng and Peng, Xin and Lou, Yiling},
  booktitle={Proceedings of the IEEE/ACM 46th International Conference on Software Engineering (ICSE)},
  pages={1--13},
  year={2024}
}

@inproceedings{liang2024large,
  title={A large-scale survey on the usability of ai programming assistants: Successes and challenges},
  author={Liang, Jenny T and Yang, Chenyang and Myers, Brad A},
  booktitle={Proceedings of the 46th IEEE/ACM International Conference on Software Engineering (ICSE)},
  pages={1--13},
  year={2024}
}

@inproceedings{piya2024llm4tdd,
  title={LLM4TDD: Best practices for test driven development using large language models},
  author={Piya, Sanyogita and Sullivan, Allison},
  booktitle={Proceedings of the 1st International Workshop on Large Language Models for Code},
  pages={14--21},
  year={2024}
}

@inproceedings{mathews2024test,
  title={Test-Driven Development and LLM-based Code Generation},
  author={Mathews, Noble Saji and Nagappan, Meiyappan},
  booktitle={Proceedings of the 39th IEEE/ACM International Conference on Automated Software Engineering (ASE)},
  pages={1583--1594},
  year={2024}
}

@inproceedings{coutinho2024role,
  title={The role of generative ai in software development productivity: A pilot case study},
  author={Coutinho, Mariana and Marques, Lorena and Santos, Anderson and Dahia, Marcio and Fran{\c{c}}a, Cesar and de Souza Santos, Ronnie},
  booktitle={Proceedings of the 1st ACM International Conference on AI-Powered Software},
  pages={131--138},
  year={2024}
}

@article{rahe2025programming,
  title={How Do Programming Students Use Generative AI?},
  author={Rahe, Christian and Maalej, Walid},
  year={2025},
  journal={Proceedings of the ACM International Conference on the Foundations of Software Engineering (FSE)}
}

@inproceedings{lin2025soen,
  title={SOEN-101: Code Generation by Emulating Software Process Models Using Large Language Model Agents},
  author={Lin, Feng and Kim, Dong Jae and Chen, TH},
  booktitle={Proceedings of the 47th IEEE/ACM International Conference on Software Engineering (ICSE)},
  year={2025}
}

@article{du2023classeval,
  title={Classeval: A manually-crafted benchmark for evaluating llms on class-level code generation},
  author={Du, Xueying and Liu, Mingwei and Wang, Kaixin and Wang, Hanlin and Liu, Junwei and Chen, Yixuan and Feng, Jiayi and Sha, Chaofeng and Peng, Xin and Lou, Yiling},
  journal={arXiv preprint arXiv:2308.01861},
  year={2023}
}

@article{ziegler2024measuring,
  title={Measuring GitHub Copilot's Impact on Productivity},
  author={Ziegler, Albert and Kalliamvakou, Eirini and Li, X Alice and Rice, Andrew and Rifkin, Devon and Simister, Shawn and Sittampalam, Ganesh and Aftandilian, Edward},
  journal={Communications of the ACM},
  volume={67},
  number={3},
  pages={54--63},
  year={2024},
  publisher={ACM New York, NY, USA}
}

@article{brown2020language,
  title={Language models are few-shot learners},
  author={Brown, Tom and Mann, Benjamin and Ryder, Nick and Subbiah, Melanie and Kaplan, Jared D and Dhariwal, Prafulla and Neelakantan, Arvind and Shyam, Pranav and Sastry, Girish and Askell, Amanda and others},
  journal={Advances in neural information processing systems (NeurIPS)},
  volume={33},
  pages={1877--1901},
  year={2020}
}

@article{kojima2022large,
  title={Large language models are zero-shot reasoners},
  author={Kojima, Takeshi and Gu, Shixiang Shane and Reid, Machel and Matsuo, Yutaka and Iwasawa, Yusuke},
  journal={Advances in neural information processing systems (NeurIPS)},
  volume={35},
  pages={22199--22213},
  year={2022}
}

@inproceedings{kang2023large,
  title={Large language models are few-shot testers: Exploring llm-based general bug reproduction},
  author={Kang, Sungmin and Yoon, Juyeon and Yoo, Shin},
  booktitle={2023 IEEE/ACM 45th International Conference on Software Engineering (ICSE)},
  pages={2312--2323},
  year={2023},
  organization={IEEE}
}

@article{yuan2023no,
  title={No more manual tests? evaluating and improving chatgpt for unit test generation},
  author={Yuan, Zhiqiang and Lou, Yiling and Liu, Mingwei and Ding, Shiji and Wang, Kaixin and Chen, Yixuan and Peng, Xin},
  journal={arXiv preprint arXiv:2305.04207},
  year={2023}
}

@article{ruiz2024novel,
  title={A Novel Approach for Automatic Program Repair using Round-Trip Translation with Large Language Models},
  author={Ruiz, Fernando Vallecillos and Grishina, Anastasiia and Hort, Max and Moonen, Leon},
  journal={arXiv preprint arXiv:2401.07994},
  year={2024}
}

@article{barke2023grounded,
  title={Grounded copilot: How programmers interact with code-generating models},
  author={Barke, Shraddha and James, Michael B and Polikarpova, Nadia},
  journal={Proceedings of the ACM on Programming Languages},
  volume={7},
  number={OOPSLA1},
  pages={85--111},
  year={2023},
  publisher={ACM New York, NY, USA}
}

@article{white2023prompt,
  title={A prompt pattern catalog to enhance prompt engineering with chatgpt},
  author={White, Jules and Fu, Quchen and Hays, Sam and Sandborn, Michael and Olea, Carlos and Gilbert, Henry and Elnashar, Ashraf and Spencer-Smith, Jesse and Schmidt, Douglas C},
  journal={arXiv preprint arXiv:2302.11382},
  year={2023}
}

@inproceedings{liu2022design,
  title={Design guidelines for prompt engineering text-to-image generative models},
  author={Liu, Vivian and Chilton, Lydia B},
  booktitle={Proceedings of the 2022 CHI conference on human factors in computing systems},
  pages={1--23},
  year={2022}
}

@inproceedings{arora2022ask,
  title={Ask me anything: A simple strategy for prompting language models},
  author={Arora, Simran and Narayan, Avanika and Chen, Mayee F and Orr, Laurel and Guha, Neel and Bhatia, Kush and Chami, Ines and Sala, Frederic and R{\'e}, Christopher},
  booktitle={International Conference on Learning Representations (ICLR)},
  year={2023}
}

@online{Upwork,
  author =    {Upwork Global Inc},
  title =  {FreeLancer Project Platform Service},
  url =    {https://www.upwork.com/},
  lastaccessed = {Feb 19 2025}
}

@online{LeetCode,
  author =    {LeetCode},
  title =  {Online Programming Learning Service},
  url =    {https://leetcode.com/problemset/},
  lastaccessed = {Feb 20 2025}
}

@online{HackerRank,
  author =    {HackerRank},
  title =  {Online Coding and Technical Assessments},
  url =    {https://www.hackerrank.com/},
  lastaccessed = {Feb 20 2025}
}

@book{burnstein2006practical,
  title={Practical software testing: a process-oriented approach},
  author={Burnstein, Ilene},
  year={2006},
  publisher={Springer Science \& Business Media}
}

@article{boehm1987improving,
  title={Improving software productivity},
  author={Boehm, Barry W.},
  journal={Computer},
  volume={20},
  number={09},
  pages={43--57},
  year={1987},
  publisher={IEEE Computer Society}
}

@incollection{scacchi1995understanding,
  title={Understanding software productivity},
  author={Scacchi, Walt},
  booktitle={Software Engineering and Knowledge Engineering: Trends for the next decade},
  pages={273--316},
  year={1995},
  publisher={World Scientific}
}

@article{mccabe1976complexity,
  title={A complexity measure},
  author={McCabe, Thomas J},
  journal={IEEE Transactions on software Engineering},
  number={4},
  pages={308--320},
  year={1976},
  publisher={IEEE}
}

@misc{ISO25002:2024,
  author       = {{International Organization for Standardization}},
  title        = {{ISO/IEC 25002:2024 --- Systems and software engineering --- Systems and software Quality Requirements and Evaluation (SQuaRE) --- Quality model for AI-based systems}},
  howpublished = {\url{https://www.iso.org/standard/78175.html}},
  year         = {2024},
  lastaccessed = {Feb 24 2025}
}

@article{fakhoury2024llm,
  title={Llm-based test-driven interactive code generation: User study and empirical evaluation},
  author={Fakhoury, Sarah and Naik, Aaditya and Sakkas, Georgios and Chakraborty, Saikat and Lahiri, Shuvendu K},
  journal={IEEE Transactions on Software Engineering},
  year={2024},
  publisher={IEEE}
}

@article{tian2023chatgpt,
  title={Is ChatGPT the ultimate programming assistant--how far is it?},
  author={Tian, Haoye and Lu, Weiqi and Li, Tsz On and Tang, Xunzhu and Cheung, Shing-Chi and Klein, Jacques and Bissyand{\'e}, Tegawend{\'e} F},
  journal={arXiv preprint arXiv:2304.11938},
  year={2023}
}

@incollection{wilcoxon1992individual,
  title={Individual comparisons by ranking methods},
  author={Wilcoxon, Frank},
  booktitle={Breakthroughs in statistics: Methodology and distribution},
  pages={196--202},
  year={1992},
  publisher={Springer}
}

@article{fisher1922interpretation,
  title={On the interpretation of $\chi$ 2 from contingency tables, and the calculation of P},
  author={Fisher, Ronald A},
  journal={Journal of the royal statistical society},
  volume={85},
  number={1},
  pages={87--94},
  year={1922},
  publisher={JSTOR}
}

@book{cramer1999mathematical,
  title={Mathematical methods of statistics},
  author={Cram{\'e}r, Harald},
  volume={9},
  year={1999},
  publisher={Princeton university press}
}

@article{cliff1993dominance,
  title={Dominance statistics: Ordinal analyses to answer ordinal questions.},
  author={Cliff, Norman},
  journal={Psychological bulletin},
  volume={114},
  number={3},
  pages={494},
  year={1993},
  publisher={American Psychological Association}
}

@article{spearman1961proof,
  title={The proof and measurement of association between two things.},
  author={Spearman, Charles},
  year={1961},
  publisher={Appleton-Century-Crofts}
}

@inproceedings{paul2024benchmarks,
  title={Benchmarks and metrics for evaluations of code generation: A critical review},
  author={Paul, Debalina Ghosh and Zhu, Hong and Bayley, Ian},
  booktitle={2024 IEEE International Conference on Artificial Intelligence Testing (AITest)},
  pages={87--94},
  year={2024},
  organization={IEEE}
}

@article{liu2024refining,
  title={Refining chatgpt-generated code: Characterizing and mitigating code quality issues},
  author={Liu, Yue and Le-Cong, Thanh and Widyasari, Ratnadira and Tantithamthavorn, Chakkrit and Li, Li and Le, Xuan-Bach D and Lo, David},
  journal={ACM Transactions on Software Engineering and Methodology},
  volume={33},
  number={5},
  pages={1--26},
  year={2024},
  publisher={ACM New York, NY}
}

@inproceedings{liang2024usability,
  author    = {Jenny T. Liang and Chenyang Yang and Brad A. Myers},
  title     = {A Large-Scale Survey on the Usability of {AI} Programming Assistants: Successes and Challenges},
  booktitle = {Proceedings of the 46th {IEEE/ACM} International Conference on Software Engineering ({ICSE} 2024)},
  pages     = {52:1--52:13},
  publisher = {{ACM}},
  year      = {2024},
  doi       = {10.1145/3597503.3608128},
  url       = {https://doi.org/10.1145/3597503.3608128}
}

@article{sergeyuk2025practice,
  author    = {Agnia Sergeyuk and Yaroslav Golubev and Timofey Bryksin and Iftekhar Ahmed},
  title     = {Using {AI}-based coding assistants in practice: State of affairs, perceptions, and ways forward},
  journal   = {Information and Software Technology},
  volume    = {178},
  pages     = {107610},
  year      = {2025},
  doi       = {10.1016/j.infsof.2024.107610},
  url       = {https://doi.org/10.1016/j.infsof.2024.107610}
}

@inproceedings{weisz2024enterprise,
  author    = {Justin D. Weisz and Shraddha Vijay Kumar and Michael J. Muller and Karen-Ellen Browne and Arielle Goldberg and Katrin Ellice Heintze and Shagun Bajpai},
  title     = {Examining the Use and Impact of an {AI} Code Assistant on Developer Productivity and Experience in the Enterprise},
  booktitle = {Extended Abstracts of the {CHI} Conference on Human Factors in Computing Systems ({CHI EA} '25)},
  articleno = {673},
  pages     = {673:1--673:13},
  publisher = {{ACM}},
  year      = {2025},
  doi       = {10.1145/3706599.3706670},
  url       = {https://doi.org/10.1145/3706599.3706670},
  eprint    = {2412.06603},
  archivePrefix = {arXiv},
  primaryClass  = {cs.SE}
}

@inproceedings{stray2025humanai,
  author    = {Viktoria Stray and Astri Barbala and Viggo Tellefsen Wivestad},
  title     = {Human-{AI} Collaboration in Software Development: A Mixed-Methods Study of Developers' Use of {GitHub Copilot} and {ChatGPT}},
  booktitle = {{FSE Companion} '25: Proceedings of the 33rd {ACM} International Conference on the Foundations of Software Engineering},
  pages     = {1325--1332},
  publisher = {{ACM}},
  year      = {2025},
  doi       = {10.1145/3696630.3730566},
  url       = {https://doi.org/10.1145/3696630.3730566}
}

@article{liang2025promptprogramming,
  author  = {Jenny T. Liang and Melissa Lin and Nikitha Rao and Brad A. Myers},
  title   = {Prompts Are Programs Too! Understanding How Developers Build Software Containing Prompts},
  journal = {Proceedings of the {ACM} on Software Engineering},
  volume  = {2},
  number  = {{FSE}},
  pages   = {1591--1614},
  year    = {2025},
  doi     = {10.1145/3729342},
  url     = {https://doi.org/10.1145/3729342}
}

@inproceedings{treude2025taxonomy,
  author    = {Christoph Treude and Marco Aur{\'e}lio Gerosa},
  title     = {How Developers Interact with {AI}: A Taxonomy of Human-{AI} Collaboration in Software Engineering},
  booktitle = {Proceedings of the 2025 {IEEE/ACM} Second International Conference on {AI} Foundation Models and Software Engineering ({Forge} 2025)},
  pages     = {236--240},
  publisher = {{IEEE}},
  year      = {2025},
  doi       = {10.1109/Forge66646.2025.00033},
  url       = {https://doi.org/10.1109/Forge66646.2025.00033}
}

@article{sergeyuk2026hax,
  author  = {Agnia Sergeyuk and Ilya Zakharov and Ekaterina Koshchenko and Maliheh Izadi},
  title   = {Human-{AI} Experience in Integrated Development Environments: A Systematic Literature Review},
  journal = {Empirical Software Engineering},
  volume  = {31},
  number  = {3},
  articleno = {55},
  year    = {2026},
  doi     = {10.1007/s10664-025-10793-0},
  url     = {https://doi.org/10.1007/s10664-025-10793-0}
}

@article{forsgren2021space,
  author  = {Nicole Forsgren and Margaret-Anne D. Storey and Chandra Shekhar Maddila and Thomas Zimmermann and Brian Houck and Jenna L. Butler},
  title   = {The {SPACE} of Developer Productivity: There's More to It Than You Think},
  journal = {{ACM Queue}},
  volume  = {19},
  number  = {1},
  pages   = {20--48},
  year    = {2021},
  doi     = {10.1145/3454122.3454124},
  url     = {https://doi.org/10.1145/3454122.3454124}
}

@article{noda2023devex,
  author  = {Abi Noda and Margaret-Anne D. Storey and Nicole Forsgren and Michaela Greiler},
  title   = {{DevEx}: What Actually Drives Productivity: The Developer-Centric Approach to Measuring and Improving Productivity},
  journal = {{ACM Queue}},
  volume  = {21},
  number  = {2},
  pages   = {35--53},
  year    = {2023},
  doi     = {10.1145/3595878},
  url     = {https://doi.org/10.1145/3595878}
}

@article{hoda2023augmentedagile,
  author  = {Rashina Hoda and Hoa Khanh Dam and Chakkrit Tantithamthavorn and Patanamon Thongtanunam and Margaret-Anne D. Storey},
  title   = {Augmented Agile: Human-Centered {AI}-Assisted Software Management},
  journal = {{IEEE Software}},
  volume  = {40},
  number  = {4},
  pages   = {106--109},
  year    = {2023},
  doi     = {10.1109/MS.2023.3268725},
  url     = {https://doi.org/10.1109/MS.2023.3268725}
}

@article{copenhagen2024manifesto,
  author  = {Daniel Russo and Sebastian Baltes and {van Berkel}, Niels and Paris Avgeriou and Fabio Calefato and Beatriz Cabrero-Daniel and Gemma Catolino and J{\"u}rgen Cito and Neil Ernst and Thomas Fritz and Hideaki Hata and Reid Holmes and Maliheh Izadi and Foutse Khomh and Kj{\ae}rgaard, Mikkel Baun and Grischa Liebel and Lafuente, Alberto Lluch and Stefano Lambiase and Walid Maalej and Gail Murphy and Moe, Nils Brede and Gabrielle O'Brien and Elda Paja and Mauro Pezz{\`e} and Persson, John Stouby and Rafael Prikladnicki and Paul Ralph and Martin Robillard and Silva, Thiago Rocha and Stol, Klaas Jan and Margaret-Anne Storey and Viktoria Stray and Paolo Tell and Christoph Treude and Bogdan Vasilescu},
  title   = {Generative {AI} in Software Engineering Must Be Human-Centered: The Copenhagen Manifesto},
  journal = {Journal of Systems and Software},
  volume  = {216},
  pages   = {112115},
  year    = {2024},
  doi     = {10.1016/j.jss.2024.112115},
  url     = {https://doi.org/10.1016/j.jss.2024.112115}
}

@inproceedings{ahn2024trust,
  author    = {Daehwan Ahn and Abdullah Almaatouq and Monisha Gulabani and Kartik Hosanagar},
  title     = {Impact of Model Interpretability and Outcome Feedback on Trust in {AI}},
  booktitle = {Proceedings of the {CHI} Conference on Human Factors in Computing Systems ({CHI} '24)},
  articleno = {27},
  pages     = {27:1--27:25},
  publisher = {{ACM}},
  year      = {2024},
  doi       = {10.1145/3613904.3642780},
  url       = {https://doi.org/10.1145/3613904.3642780}
}

\end{document}